\documentclass{aastex61}

\newcommand\aastex{AAS\TeX}

\newcommand\uchii{\mbox{UCH\small{II}\normalsize}}

\newcommand\amin{\mbox{$^\prime$}}%
\newcommand\adeg{\mbox{$^\circ$}}%
\newcommand{\HII}{H {\small{II}} }
\newcommand{\kms}{{\rm km~s}^{-1}}

\newcommand{\cmcub}{cm$^{-3}$}

\received{xx xx, 2017}
\revised{xx xx, 2017}
\accepted{xx xx, 2017}

\shorttitle{\aastex\  Molecular Gas and  star formation around H {\small{II}} region Sh2-104}
\shortauthors{Xu et al.}

\begin{document}

\title{Two-Dimensional Molecular Gas and ongoing star formation around H {\small{II}} region Sh2-104}

\correspondingauthor{Jin-Long Xu}
\email{xujl@bao.ac.cn}

\author[0000-0002-0786-7307]{Jin-Long Xu}
\affil{National Astronomical Observatories, Chinese Academy of Sciences, Beijing 100012, China}

\author{Ye Xu}
\affiliation{Purple Mountain Observatory, Chinese Academy of Sciences, Nanjing 210008, China}
\affiliation{Purple Mountain Observatory, Qinghai Station, 817000, Delingha, China}

\author{Naiping Yu}
\affiliation{National Astronomical Observatories, Chinese Academy of Sciences, Beijing 100012, China}

\author{Chuan-peng Zhang}
\affiliation{National Astronomical Observatories, Chinese Academy of Sciences, Beijing 100012, China}

\author{Xiao-Lan Liu}
\affiliation{National Astronomical Observatories, Chinese Academy of Sciences, Beijing 100012, China}

\author{Jun-Jie Wang}
\affiliation{National Astronomical Observatories, Chinese Academy of Sciences, Beijing 100012, China}

\author{Chang-chun Ning}
\affiliation{Tibet University, Lhasa,  Tibet 850000, China}

\author{Bing-Gang Ju}
\affiliation{Purple Mountain Observatory, Qinghai Station, 817000, Delingha, China}

\author{Guo-Yin Zhang}
\affiliation{National Astronomical Observatories, Chinese Academy of Sciences, Beijing 100012, China}

\begin{abstract}
We performed a multiwavelength study towards \HII region Sh2-104.  New maps of $^{12}$CO $J$=1-0 and $^{13}$CO $J$=1-0 were obtained from the Purple Mountain Observatory (PMO) 13.7 m radio telescope. Sh2-104 displays a double-ring structure.  The outer ring with a radius of 4.4 pc is dominated by 12 $\mu$m, 500 $\mu$m, $^{12}$CO $J$=1-0,  and $^{13}$CO $J$=1-0 emission, while the inner ring with  a radius of 2.9 pc is dominated by 22 $\mu$m and 21 cm emission.  We did not detect CO emission  inside the outer ring. The north-east portion of the outer ring is blueshifted, while the south-west portion is redshifted. The present observations have provided evidence that the collected outer ring around Sh2-104 is a two-dimensional  structure. From the column density map constructed by the Hi-GAL survey data, we extract 21 clumps.  About 90\% of all the clumps will form low-mass stars. A power-law fit to the clumps yields $M=281M_{\odot}(r/\rm pc)^{1.31\pm0.08}$.  The selected YSOs are associated with the collected material on the edge of Sh2-104.  The derived dynamical age of Sh2-104 is 1.6$\times10^{6}$ yr. Compared the Sh2-104 dynamical age with the YSOs timescale and the fragmentation time of the molecular ring, we further confirm that collect-and-collapse process operates in this region, indicating a positive feedback from a massive star for surrounding gas.

\end{abstract}


\section{Introduction} \label{sec:intro}

OB stars emit copious ultraviolet (UV) photons. The UV photons with energies above 13.6 eV can ionize hydrogen to create an \HII region. Since the gas pressure inside HII regions is higher than  that of their surrounding neutral medium, the \HII regions expand. Expanding \HII regions will also reshape their surrounding molecular gas, and thereby regulate star formation in the molecular gas. Accordingly, it will brings two aspect of roles. One is the destructive role, which can evaporate or disperse the surrounding molecular gas and consequently may terminate star formation, and  another is the constructive role, which can trigger star formation. 

At present two main processes have been put forward for triggering star formation at the peripheries of \HII regions: collect and collapse (CC) and radiation-driven implosion (RDI) \citep{elmegreen1977,deha2010}.  In the CC process, a compressed layer of neutral material is accumulated between the ionization front and shock front, and star formation occurs when this layer becomes gravitationally unstable. Unlike the CC precess, the RDI process that a pre-existing denser gas is compressed by the shocks, and then form stars. Howerver, the CC process is more attractive because it allows the formation of massive stars or clusters \citep{deha2005}. To investigate the triggered star formation by the CC process, several individual \HII regions have been well studied, such as Sh-104 \citep{deha2003}, RCW 79 \citep{Zavagno2006},  RCW 120 \citep{Zavagno2007}, Sh2-212 \citep{deha2008}, Sh2-217 \citep{Brand2011}, Sh2-90 \citep{Samal2014}, Sh2-87 \citep{Xu2014}, N6 \citep{Yuan2014} and Gum 31 \citep{Duronea2015}. Moreover, there are also some statistical studies for infrared bubbles \citep[e.g.,][] {deha2010,Kendrew2012,Thompson12,Kendrew2016}, which are created by  the expanding \HII regions. A common feature of the above studies that several massive fragments are found on the border of each \HII region.  Some star formation activities have been detected in these fragments, such as outflow, $\uchii$, and water masers.   Generally, if an expanding \HII region collect their surrounding gas, the morphology of the molecular gas will show three-dimensional spherical structure. CO observations can provide velocity information to reveal the gas structure surrounded \HII regions.  \citet{beau2010} observed 43 infrared bubbles using CO molecular line, but they did not detect the front and back faces of these shells at blueshifted and redshifted velocities. Hence, they concluded that the bubbles enclosing \HII regions are two-dimensional rings formed in parental molecular clouds with thicknesses not greater than the bubble sizes. However, the molecular gas  with a three-dimensional spherical structure or a two-dimensional ring around \HII region is important for understanding the triggered star formation \citep{deha2015}.

Sh2-104 is an optically visible Galactic \HII region with a 7$\amin$ diameter. The distance to Sh2-104 is $\sim$4 kpc \citep{deha2003}. This \HII region is excited by an O6V center star \citep{Crampton1978,Lahulla1985}, whose ionized gas has a LSR velocity of 0 $\kms$ \citep{Georgelin1973}. Sh2-104 displays a shell-like morphology, detected at optical and radio wavelengths. Using the Herschel data, \citet{Podon2010} found that the dust temperature is $\sim$25 K in the photodissociation region (PDR) of Sh2-104, while it is $\sim$40 K in its interior. \citet{deha2003} detected a molecular shell with four large molecular condensations around Sh2-104.  An $\uchii$ region  lies at an eastern condensation, which is associated with IRAS 20160+3636. High CN/HCN ratio in the eastern condensation suggests that the condensation is affected by  Sh2-104 \citep{Minh2014}. A near-IR cluster lies in the $\uchii$ direction \citep{deha2003}.  The massive condensations and cluster around  Sh2-104 suggests that this \HII region  is a typical candidate for triggering star formation by CC process.

In this paper,  we performed a multi-wavelength study to further investigate the gas structure and star formation around \HII region  Sh2-104. The molecular gas associated with Sh 104 was observed in $^{12}$CO $J$=2-1, $^{13}$CO $J$=1-0, and C$^{18}$O $J$=1-0 with the IRAM 30m telescope \citep{deha2003}. Howerver, they did not give the maps of the $^{13}$CO $J$=1-0 and C$^{18}$O $J$=1-0 molecular gas in their paper. New maps of $^{12}$CO $J$=1-0 and $^{13}$CO $J$=1-0 were shown from the Purple Mountain Observatory (PMO) 13.7m radio telescope. Combining our data with those obtained by the NRAO VLA Sky survey, the Hi-GAL survey, and the WISE survey, our aim was to construct a comprehensive large-scale picture of Sh2-104. Our observations and data reduction are described in Sect.\ref{sect:data}, and the results are presented in Sect.\ref{sect:results}. In Sect.\ref{sect:discu}, we  discuss the gas structure around Sh2-104 and star formation scenario, while our conclusions are summarized in Sect.\ref{sect:summary}.

\section{Observation and data processing}
\subsection{Archival data}
\label{sect:archive}
We used far-infrared data (70 $\mu$m $\sim$ 500 $\mu$m) from the Herschel Infrared Galactic Plane survey \citep[Hi-GAL;][]{Molinari2010} carried out by the Herschel Space Observatory. The initial survey covered a Galactic longitude region of 300$^{\circ}$$<$$\ell$$<$60$^{\circ}$ and $|b|$$<$1.0$^{\circ}$.  The Hi-GAL survey used two instruments: PACS \citep{Poglitsch2010} and SPIRE \citep{Griffin2010} in paralel mode to carry out a survey of the inner Galaxy in five bands: 70, 160, 250, 350 and 500 $\mu$m. The scan speeds of PACS is 20$^{\prime\prime}$  per second, while  30$^{\prime\prime}$ per second for SPIRE.  The angular resolutions of these five bands  are 10$\farcs7$, 11$\farcs4$, 18$\farcs2$, 24$\farcs9$, and 36$\farcs3$, respectively. These Hi-GAL data have been used to explore the flux density distribution and to construct column density map of the studied region.

The 1.4 GHz radio continuum emission data were obtained from the NRAO VLA Sky Survey  \citep[NVSS;][]{Condon1998} which is a 1.4 GHz continuum survey covering the entire sky north of -40$^{\circ}$ declination with a noise of about 0.45 mJy/beam and a resolution of 45$^{\prime\prime}$.

We also utilized infrared data from the survey of the Wide-field Infrared Survey Explorer \citep[WISE;][]{Wright2010}. The WISE survey is mapping the whole sky in four infrared bands 3.4 $\mu$m (W1), 4.6 $\mu$m (W2), 12 $\mu$m (W3), and 22 $\mu$m (W4) at angular resolutions of 6$\farcs1$, 6$\farcs4$, 6$\farcs5$, and 12$^{\prime\prime}$, respectively.

\label{sect:data}
\subsection{Purple Mountain Data}
We also made the mapping observations of \HII region Sh2-104 and its adjacent region in the transitions of $^{12}$CO $J$=1-0, $^{13}$CO $J$=1-0 and C$^{18}$O $J$=1-0 lines using the Purple Mountain Observation (PMO) 13.7 m radio telescope at De Ling Ha in the west of China at an altitude of 3200 meters, during December 2016. The 3$\times$3 beam array receiver system in single-sideband (SSB) mode was used as front end. The back end is a fast Fourier transform spectrometer (FFTS) of 16384 channels with a bandwidth of 1 GHz, corresponding to a velocity resolution of 0.16 km s$^{-1}$ for $^{12}$CO $J$=1-0, and 0.17 km s$^{-1}$ for $^{13}$CO $J$=1-0 and C$^{18}$O $J$=1-0. $^{12}$CO $J$=1-0 was observed at upper sideband with a system noise temperature (Tsys) of 272 K, while $^{13}$CO $J$=1-0 and C$^{18}$O $J$=1-0 were observed simultaneously at lower sideband with a  system noise temperature of 145 K.  The half-power beam width (HPBW) was 53$^{\prime\prime}$ at 115 GHz and the main beam efficiency was 0.5. The pointing accuracy of the telescope was better than 5$^{\prime\prime}$, which was derived from continuum observations of planets (Venus, Jupiter, and Saturn). The source W51D (19.2 K) was observed once per hour as flux calibrator. The mean rms noise level of the calibrated brightness temperature was 0.4 K for $^{12}$CO $J$=1-0, while 0.2 K for  $^{13}$CO $J$=1-0 and C$^{18}$O $J$=1-0. Mapping observations were centered at RA(J2000)=$20^{\rm h}17^{\rm m}45^{\rm s}$, DEC(J2000)=$36^{\circ}46'00^{\prime\prime}$ using the on-the-fly mode with a constant integration time of 14 second at each point. The total mapping area is $20^{\prime}\times 20^{\prime}$ in
$^{12}$CO $J$=1-0, $^{13}$CO $J$=1-0, and C$^{18}$O $J$=1-0 with a $0.5^{\prime}\times0.5^{\prime}$
grid. The standard chopper wheel calibration technique is used to measure antenna temperature $T_{\rm A} ^{\ast}$ corrected for atmospheric absorption \citep{Kutner1981}. The final data was recorded in brightness temperature scale of $T_{\rm mb}$ (K). The data were reduced using the GILDAS/CLASS \footnote{http://www.iram.fr/IRAMFR/GILDAS/} package.

\section{Results}
\label{sect:results}
\subsection{Infrared and Radio Continuum Images}

\begin{figure}
\centering
\includegraphics[width = 0.5 \textwidth]{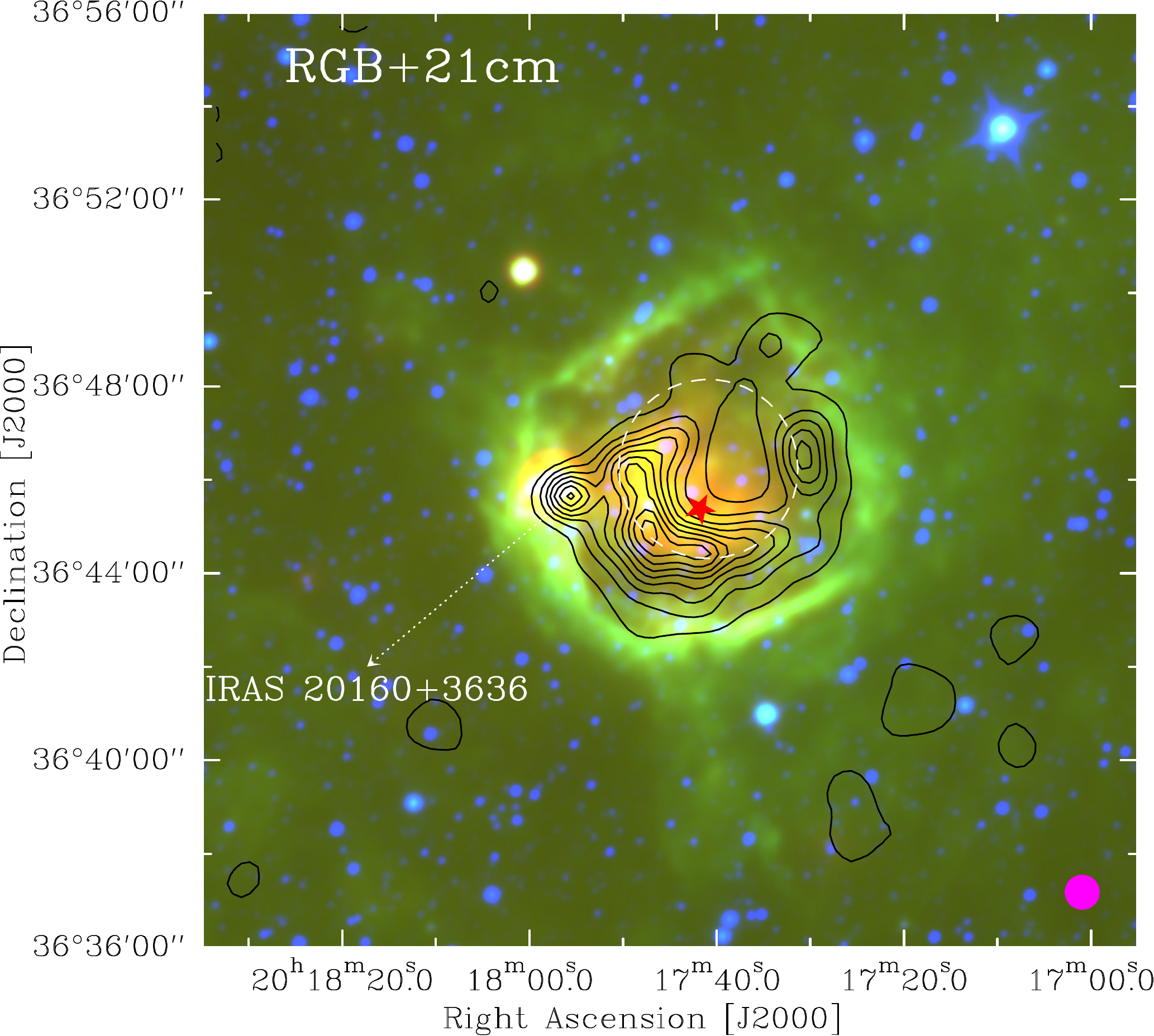}
\caption{1.4 GHz radio continuum contours in black colour, overlaid on the three color image of \HII Sh2-104 composed from the WISE 3.4 $\mu$m, 12 $\mu$m, and 22 $\mu$m bands in blue, green, and red, respectively.  The black contours begin at 5$\sigma$ in steps of 10$\sigma$, with 1$\sigma$ = 0.7 mJy beam$^{-1}$.  The white dashed circle indicates an ionized gas ring for Sh2-104. The red "star" marks the position of the exciting star.}
\label{Fig:S104-RGB}
\end{figure}

Figure \ref{Fig:S104-RGB} shows composite three-color image of Sh2-104. The three infrared bands are the WISE 3.4 $\mu$m (in blue), 12 $\mu$m (in green), and 22 $\mu$m (in red). The WISE 12 $\mu$m band contains polycyclic aromatic hydrocarbon (PAH) emission at 11.2 $\mu$m and 12.7 $\mu$m \citep{Tielens2008}. The PAH molecules are excited by the UV radiation from \HII region, but are easily destroyed inside ionized region.  Hence, the WISE 12 $\mu$m similar to the Spitzer IRAC 8.0 $\mu$m can be used to trace PDR, and  delineate \HII region boundaries \citep{Pomares2009}. In Fig. \ref{Fig:S104-RGB}, the PAH emission shows a ring-like structure with an opening towards the north. For \HII region, the WISE 22 $\mu$m emission traces heated small dust grains, which is similar to the MIPS 24 $\mu$m  band \citep{Anderson2014}. The hot dust emission in Fig. \ref{Fig:S104-RGB} displays two sources, one related to Sh2-104, and the other associated with IRAS 20160+3636. 

\begin{figure*}
\centering
\includegraphics[width = 0.48 \textwidth]{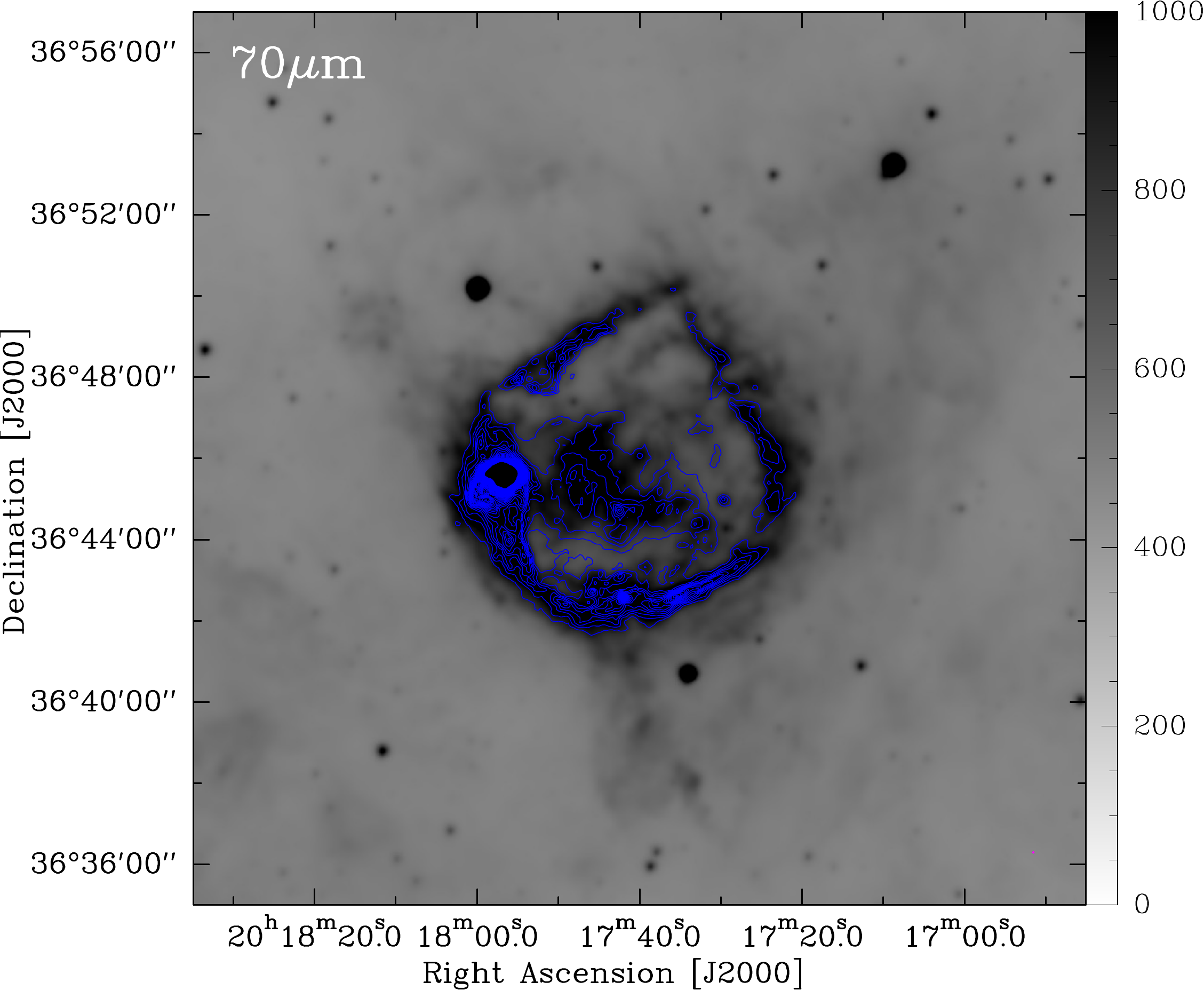}
\includegraphics[width = 0.48 \textwidth]{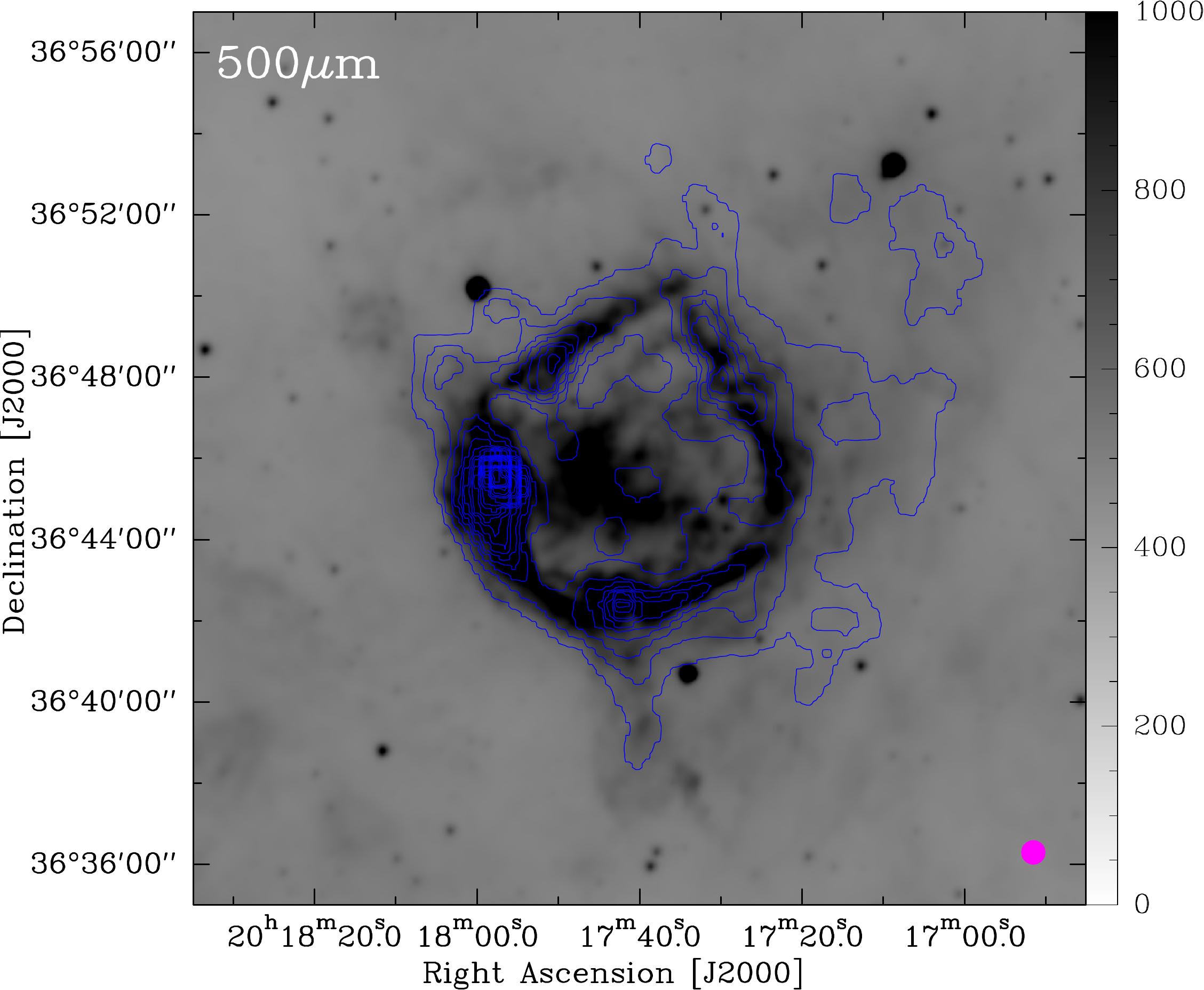}
\caption{The Herschel 70 $\mu$m and 500 $\mu$m emission  maps superimposed on the WISE 12 $\mu$m emission (grey), respectively. The units for each colour bar are Jy pixel$^{-1}$.}
\label{Fig:S104-70}
\end{figure*}

The NVSS 20 cm radio continuum emission can be used to trace ionized gas, which is also overlaid in Fig. \ref{Fig:S104-RGB} by black contours. The ionized gas emission in Fig. \ref{Fig:S104-RGB} indicates that Sh2-104 is a shell-like \HII region, whose radius is about 2.5\amin, marked by a white dashed circle.  Moreover, the ionized gas emission are spatially coincident with the hot dust emission observed at the WISE 22 $\mu$m (red color). Both the ionized gas and hot dust emission are enclosed by the PAH emission. Extending towards the east, an $\uchii$ region is located at the eastern border of Sh2-104, which is related to source IRAS 20160+3636.

The 70 $\mu$m emission is mostly  produced by relatively hot dust \citep{Faimali2012}, and partly corresponds to cool dust emission \citep{Anderson2012}.  Figure \ref{Fig:S104-70} (left panel) is the 70 $\mu$m  emission map, overlaid with the WISE 12 $\mu$m emission. The 70 $\mu$m emission also displays the ring-like shape. The ring may represent the cool dust emission, which is spatially associated with the PAH emission. Inside the ring, there is an arc-like structure. From Figures \ref{Fig:S104-RGB} and \ref{Fig:S104-70} (left), it is seen that the arc is spatially coincident with the 20 cm radio continuum emission, indicating that this part of the 70 $\mu$m emission is producted from the hot dust. The 500 $\mu$m emission originates from cool dust with an average temperature of 26 K along the PDR \citep{Anderson2012}. Figure \ref{Fig:S104-70} (right panel) shows the 500 $\mu$m  emission map superimposed on the WISE 12 $\mu$m emission. The 500 $\mu$m emission also shows the ring-like structure.  Comparing with the 12 $\mu$m and 70 $\mu$m emission ring, it becomes more extended from the 500 $\mu$m emission. In Fig.\ref{Fig:S104-70} (right panel),  the cool dust is mainly concentrated in  four large dust  clumps, which are regularly distributed at the ring-like structure around Sh2-104.

\begin{figure*}
\centering
\includegraphics[width = 0.48 \textwidth]{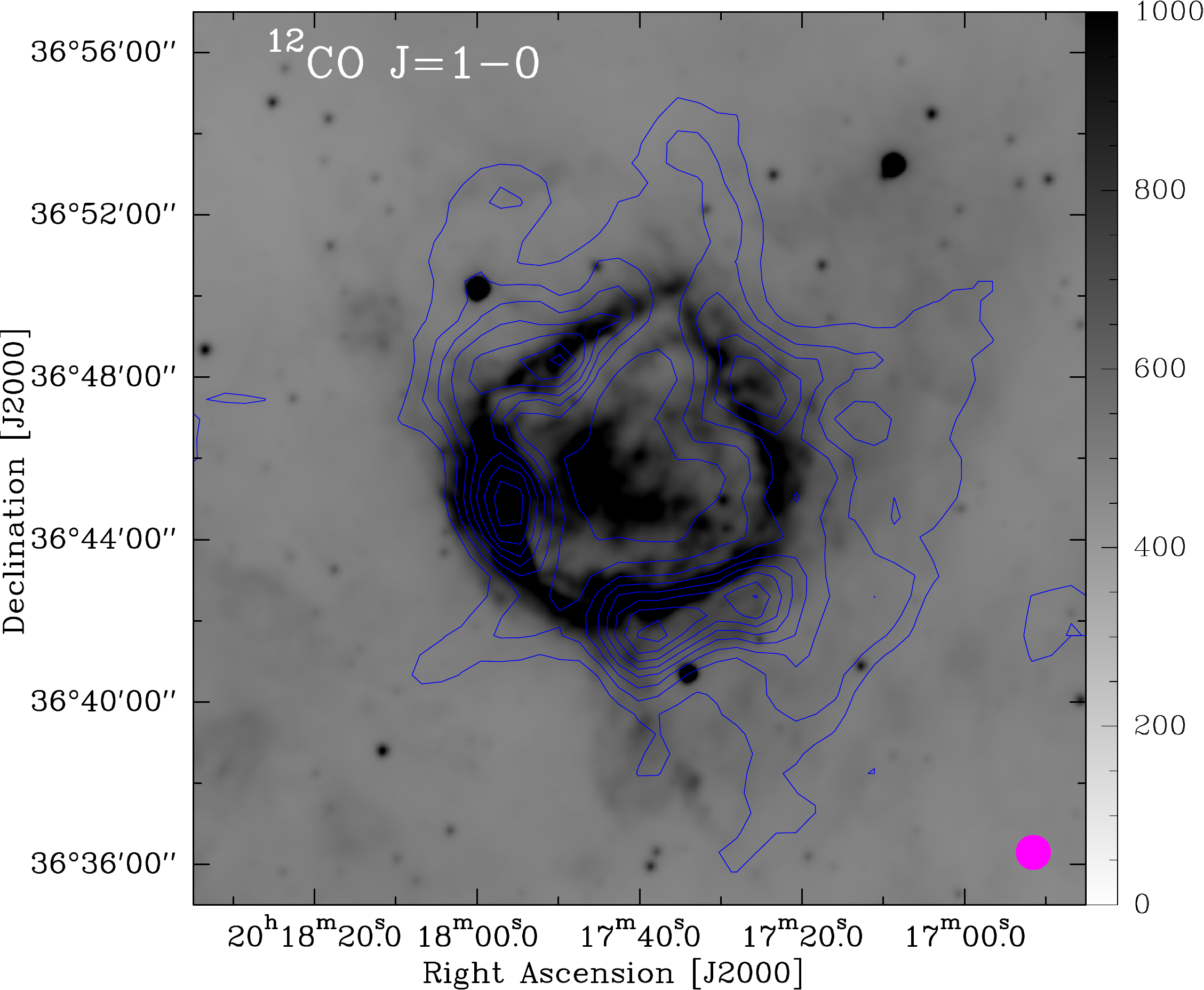}
\includegraphics[width = 0.48 \textwidth]{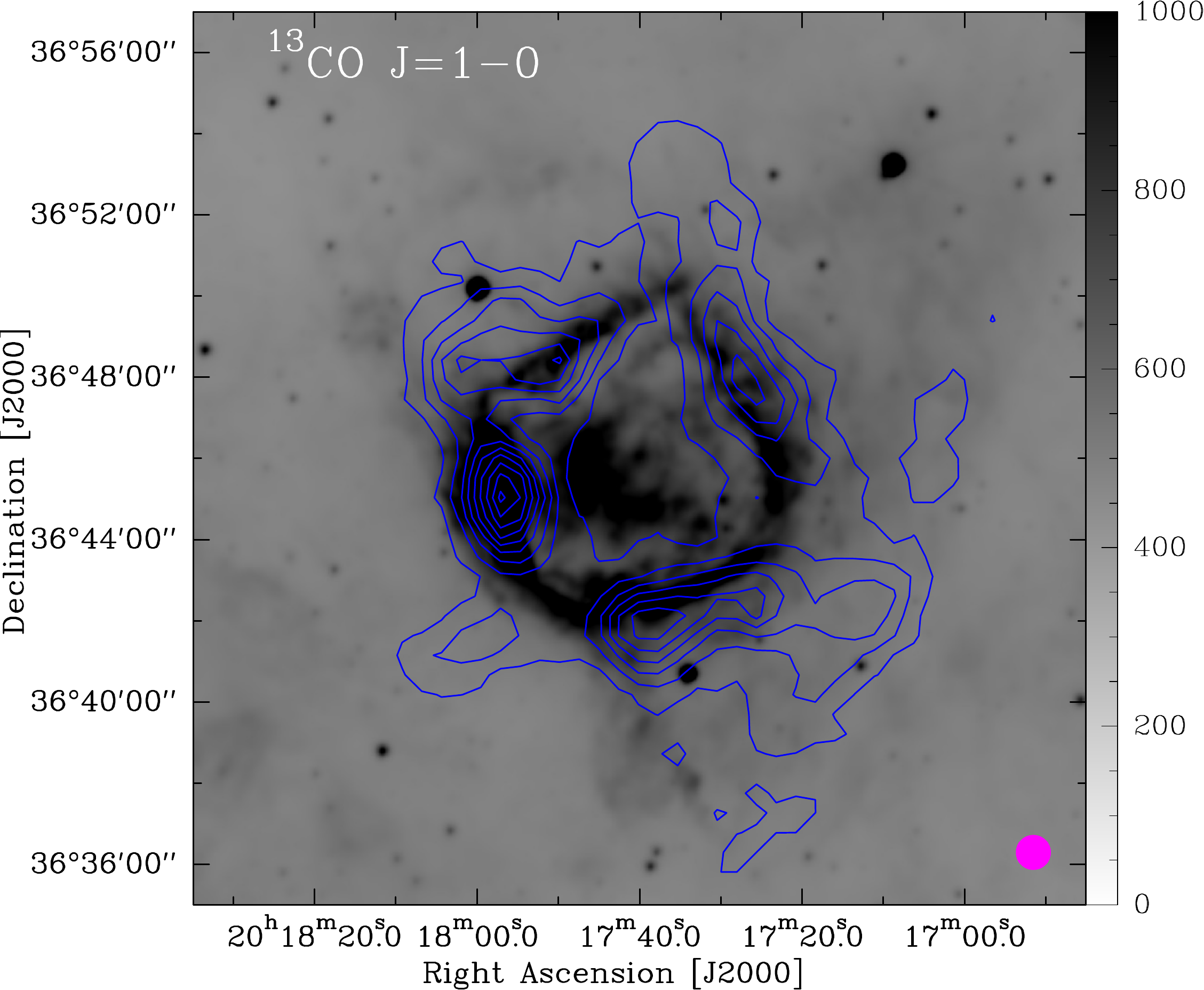} 
\caption{Left panel: Contours of $^{12}$CO $J$=1-0 emission superimposed on the 12 $\mu$m emission map (grey). The integrated velocity is from -3.4 to 4.2 km s$^{-1}$. The blue contour levels are from 14.7 (3$\sigma$) to 66.2  by a step of 7.3 K km s$^{-1}$ . Right panel: Contours of $^{13}$CO $J$=1-0 emission superimposed on the 8 $\mu$m emission map (grey). The blue contour levels are from 1.8 (3$\sigma$) to 16.2  by a step of 1.8 K km s$^{-1}$. }
\label{Fig:S104-12CO}
\end{figure*}

\subsection{CO Molecular Emission}
\begin{figure}
\vspace{4mm}
\centering
\includegraphics[width = 0.55 \textwidth]{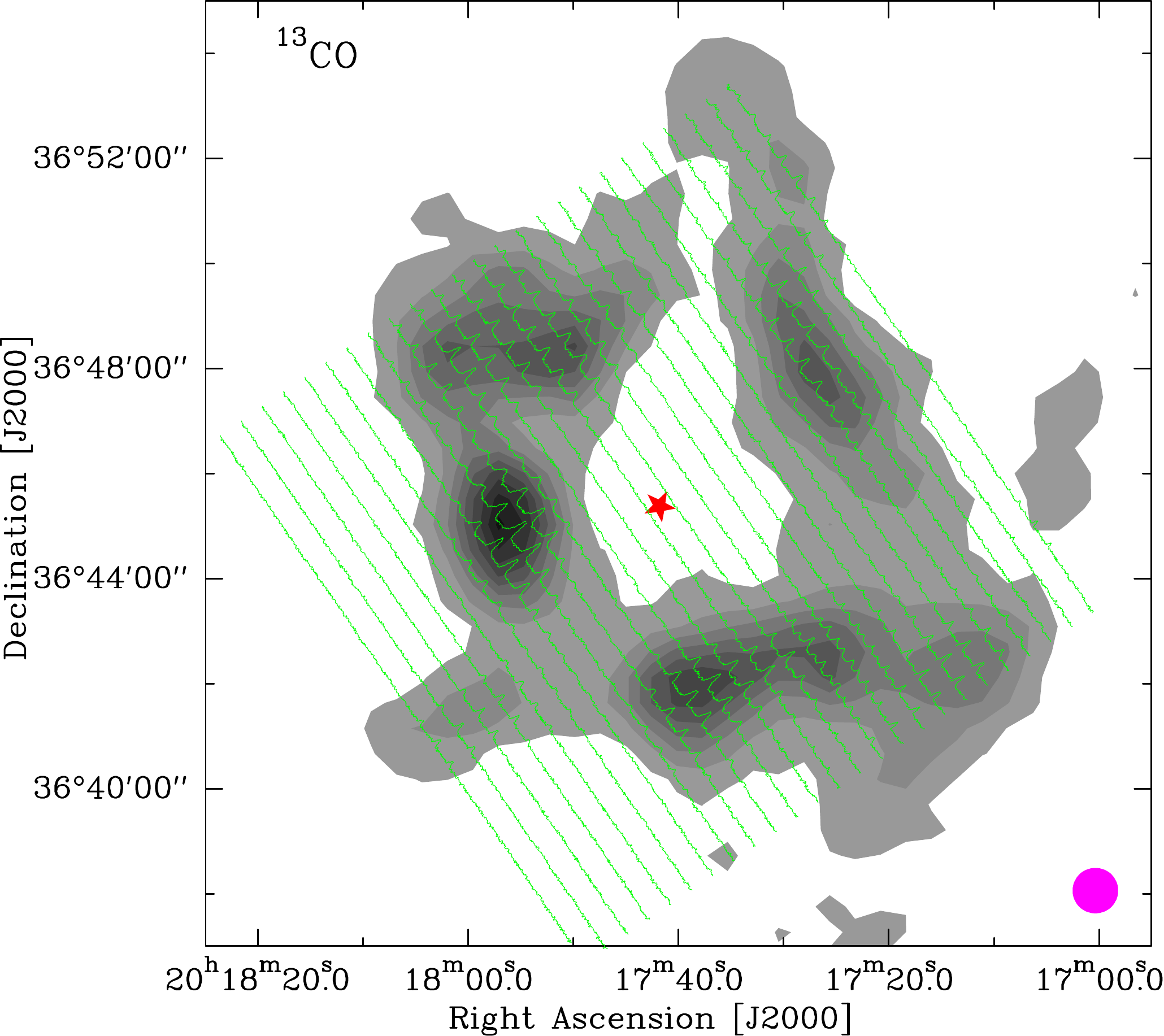}
\caption{$^{13}$CO $J$=1-0 spectra overlaid in the integrated intensity image of $^{13}$CO $J$=1-0 .}
\label{Fig:13CO-spectra}
\end{figure}

Compared with dust continuum  emission, CO emission with velocity information  can be used to separate whether the projected molecular gas is associated with  \HII region.  To analyze the morphology of molecular gas consistent with \HII region Sh2-104, here we use $^{12}$CO $J$=1-0, $^{13}$CO $J$=1-0, and C$^{18}$O $J$=1-0 lines to trace the molecular gas.  The C$^{18}$O $J$=1-0 emission was detected in some points with strong $^{13}$CO $J$=1-0 emission, suggesting that these positions are more dense. Since the C$^{18}$O $J$=1-0 signal is so week that we do not examine its spatial distribution. Comparing the optically thick $^{12}$CO line, the $^{13}$CO line is more suited to trace relatively dense gas. Using channel maps of $^{13}$CO line, we inspect gas component for our studied region.  \citet{Georgelin1973} found that the LSR velocity of the ionized gas for Sh2-104 is about 0 $\kms$.  The $^{12}$CO $J$=1-0 and $^{13}$CO $J$=1-0 emission from -3.4 to 4.2 $\kms$ is therefore found to be associated with  Sh2-104.   Figure \ref{Fig:S104-12CO} shows the integrated intensity images of $^{12}$CO $J$=1-0 and $^{13}$CO $J$=1-0, integrated over the velocity range from -3.4 to 4.2 $\kms$.  Both the $^{12}$CO $J$=1-0 and $^{13}$CO $J$=1-0 emission exhibit a ring-like shape, but the ring traced by $^{12}$CO $J$=1-0 is diffuse than that of $^{13}$CO $J$=1-0. In addition, in Fig. \ref{Fig:S104-12CO}, the $^{12}$CO $J$=1-0 and $^{13}$CO $J$=1-0 emission coincides well with  the PDR traced by the PAHs emission surrounded Sh2-104.  It is apparent that the PDR is thinner compared to the CO isotopologues emission. Inside the ring, we did not detect the stronger CO emission, which is similar to that observed in $^{12}$CO $J$=2-1 \citep{deha2003}. Figure \ref{Fig:13CO-spectra} shows the $^{13}$CO $J$=1-0 spectra on top of its integrated intensity map. It is obvious that there is not higher than the baseline signal inside the ring, which is different from that  for \HII region RCW 120. Inside RCW 120, \citet{Anderson2015} detected multiple velocity components in $^{13}$CO $J$=1-0 .

\begin{figure}
\vspace{4mm}
\centering
\includegraphics[width = 0.55 \textwidth]{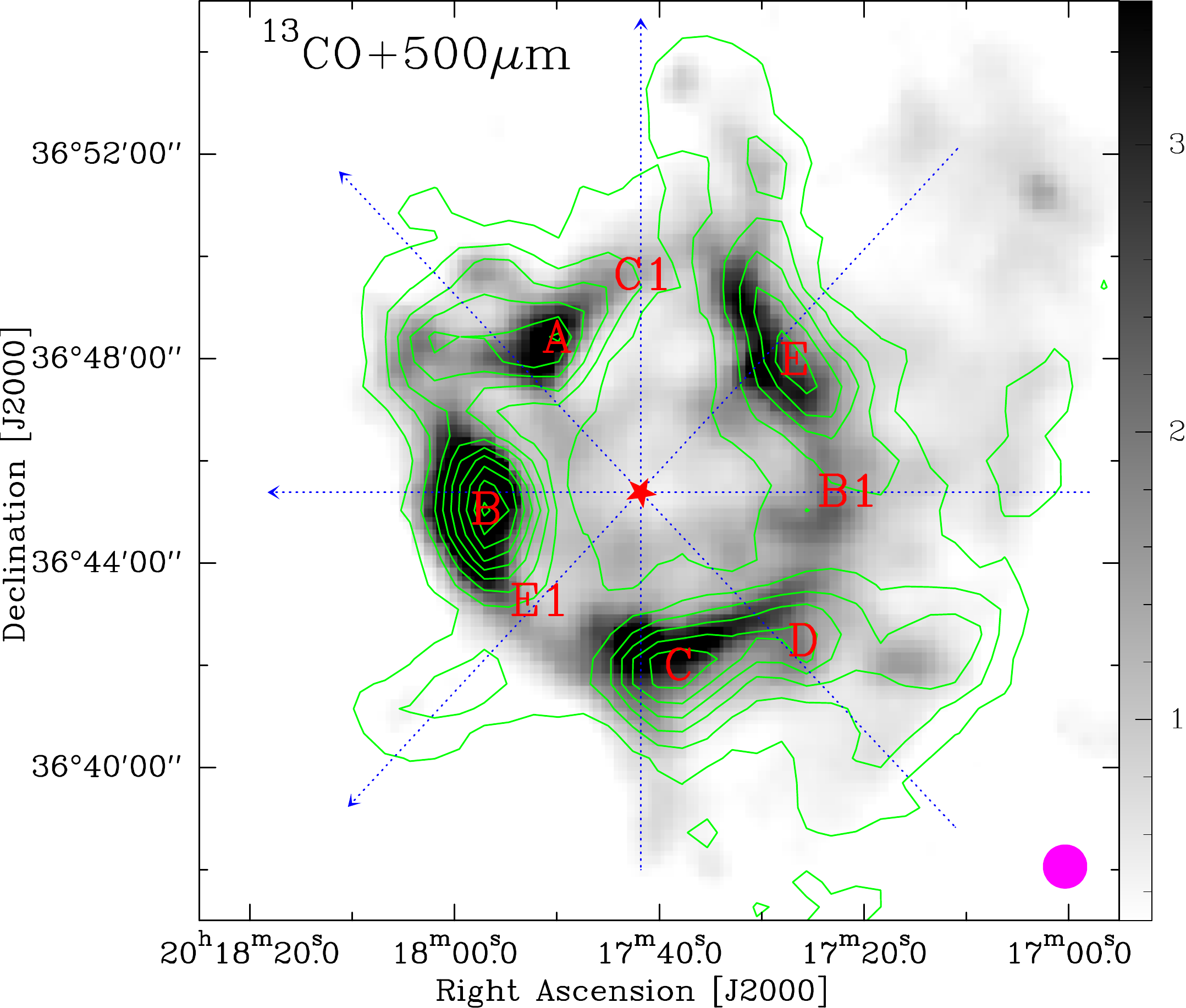}
\caption{$^{13}$CO $J$=1-0 integrated intensity map overlaid on the  Herschel 500 $\mu$m image (color scale). The green contour levels are from 1.8 (3$\sigma$) to 16.2  by a step of 1.8 K km s$^{-1}$.  Letters A, B, C, and D
indicate the different molecular clumps, while letters A1, B2, C1, and D1 represent the symmetrical position with these clumps in the ring. The red "star"  represent the exciting star. The data used in the position-velocity diagrams (in Figure 6) are selected along the blue arrows. The unit for the colour bar is Jy pixel$^{-1}$.}
\label{Fig:S104-13CO-500}
\end{figure}

\begin{figure}
\centering
\includegraphics[width = 0.46 \textwidth]{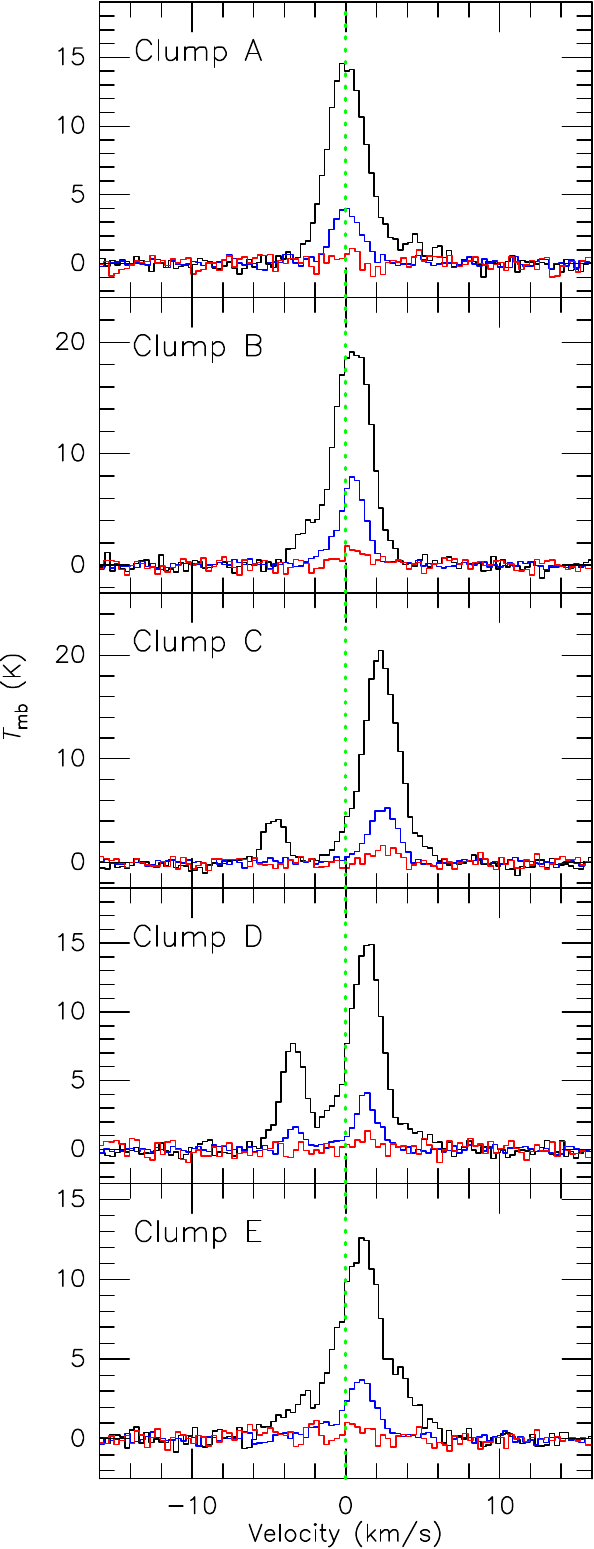}
\caption{$^{12}$CO $J$=1-0, $^{13}$CO $J$=1-0, and C$^{18}$O $J$=1-0 spectra at the peak position of the molecular clumps A, B, C, D, and E. The green dashed line indicates the systemic velocity (0 km s$^{-1}$) of Sh2-104.}
\label{Fig:S104-spectra}
\end{figure}

To investigate spatial correlation between gas and dust of the ring around Sh2-104, we made the $^{13}$CO $J$=1-0 integrated intensity map overlaid on the  Herschel 500 $\mu$m emission (Fig. \ref{Fig:S104-13CO-500}). The $^{13}$CO $J$=1-0 emission in the Fig. \ref{Fig:S104-13CO-500} is associated with the 500 $\mu$m dust emission in morphology. There are four gas fragments located in the ring. It is different from dust fragments, the southern gas fragment contains two clumps.  We designated these clumps at clump A, clump B, clump C, clump D, and clump E.
The clump B is more compact relative to other clumps, which is spatially consistent with IRAS 20160+3636.  Figure \ref{Fig:S104-spectra} shows the $^{12}$CO $J$=1-0, $^{13}$CO $J$=1-0, and C$^{18}$O $J$=1-0 spectra toward the peak position of each clump. The line profiles of $^{12}$CO $J$=1-0 and $^{13}$CO $J$=1-0 in clump B appear to only show the blue wings, while it displays the blue and red winds in clump E.  For clumps C and D, the line profiles of $^{12}$CO $J$=1-0 and $^{13}$CO $J$=1-0 show two velocity components. We fitted each spectrum with the Gaussian profile. Table \ref{Table:clump} shows the fitted results, including their peak intensities,  FWHMs, and  central velocities. From Table \ref{Table:clump}, we can see that the central velocities of these clumps are different. As shown in the Fig. \ref{Fig:S104-spectra},  the central velocity of each clump has a shift  with respect to the LSR velocity of the ionized gas for Sh2-104, except for the clump A.  

\subsection{Column Density and Mass of Molecular Ring}
To estimate the column density and mass of the molecular ring around Sh2-104, we used the optical thin $^{13}$CO $J$=1-0 emission. Assuming local thermodynamical equilibrium (LTE), the column density was estimated via Garden et al. \citep{Garden}
\begin{equation}
\mathit{N(\rm ^{13}CO)}=4.71\times10^{13}\frac{T_{\rm ex}+0.88}{\rm exp(-5.29/T_{\rm ex})}\frac{\tau}{\rm 1-exp(-\tau)}W ~\rm cm^{-2},
\end{equation}
where $\tau$ is the optical depth and $T_{\rm ex}$ is the mean excitation temperature of the molecular gas. W is $\int T_{\rm mb}dv$ in units of K km s$^{-1}$, $T_{\rm mb}$ is the corrected main-beam temperature of $^{13}$CO $J$=1-0 and $dv$ is the velocity range.

\begin{table}
\begin{center}
\tabcolsep 0.2mm\caption{Observed parameters of each line for each clump}
\def\temptablewidth{8\textwidth}%
\vspace{-2mm}
\begin{tabular}{llcccccccr}
\hline\hline\noalign{\smallskip}
Name   &&  $^{12}$CO J=1-0   &   &&&   & $^{13}$CO J=1-0 &  & \\
\cline{2-4}\cline{7-10}
        &   $T_{\rm mb}$   &FWHM  &$V_{\rm LSR}$   &&&   $T_{\rm mb}$   &FWHM  &$V_{\rm LSR}$   \\
        &  (K)               &(km $\rm s^{-1}$) &(km $\rm s^{-1}$) &&&   (K)   &(km $\rm s^{-1}$)  &(km $\rm s^{-1}$) \\
\cline{2-4}\cline{7-10}
A      & 14.4 & 3.1 & 0.1   &&&  4.0& 2.0 & 0.1   \\  
B      & 19.6 & 2.9 & 0.5   &&&  7.7& 2.0 & 0.6   \\  
C      & 19.5 & 2.8 & 2.3   &&&  5.3& 2.2 & 2.4   \\  
D      & 14.3 & 2.7 & 1.3   &&&  3.8& 1.8 & 1.3   \\  
E      & 11.0 & 4.0 & 1.0   &&&  3.5& 2.6 & 1.0   \\  
\noalign{\smallskip}\hline
\end{tabular}\end{center}
\label{Table:clump}
\end{table}

Generally, the $^{12}$CO  emission is optical thick, so we used $^{12}$CO  to estimate $T_{\rm ex}$ via  following the equation  \citep{Garden}
\begin{equation}
\mathit{T_{\rm ex}}=\frac{5.53}{{\ln[1+5.53/(T_{\rm mb}+0.82)]}},
\end{equation}
where $T_{\rm mb}$ is the corrected main-beam brightness temperature of $^{12}$CO. From this equation, we derived that the excitation temperature of the five clumps is from 14.4 K to 23.1 K. Since these clumps are regularly located in the ring, then we adopt the mean excitation temperature of about 19.2 K as that of the molecular ring.

Moreover, we assumed that the excitation temperatures of $^{12}$CO and $^{13}$CO have the same value in the molecular ring.  The optical depth ($\tau$) can be derived using the following equation  \citep{Garden}
\begin{equation}
\mathit{\tau(\rm ^{13}CO)}=-\ln[1-\frac{T_{\rm mb}}{5.29/[\rm exp(5.29/\it T_{\rm ex})\rm -1]-0.89}],
\end{equation}
The obtained optical depth is 0.25-0.67,  suggesting that the $^{13}$CO emission is optically thin in the ring.

In addition, we used the relation $N(\rm H_{2})/\it N(\rm ^{13}CO)$ $\approx$
$7\times10^{5}$  \citep{Castets} to estimate the H$_{2}$ column density. The mass of the ring can be determined by
\begin{equation}
\mathit{M_{\rm H_{2}}}=\mu m_{\rm H}\it N(\rm H_{2})\it S,
\end{equation}
where $\mu$=2.72 is the mean molecular weight, $m_{\rm
H}$ is the mass of a H atom, and $S$ is the projected 2D area of the ring. Hence, we obtained that the mean column density of the ring is 6.8$\times10^{21}$ cm$^{-2}$. Using the mean column density, we derived a mass of  $\sim$2.2$\times10^{4}\rm M_{ \odot}$ for the ring.

\begin{figure}
\centering
\includegraphics[width = 0.7 \textwidth]{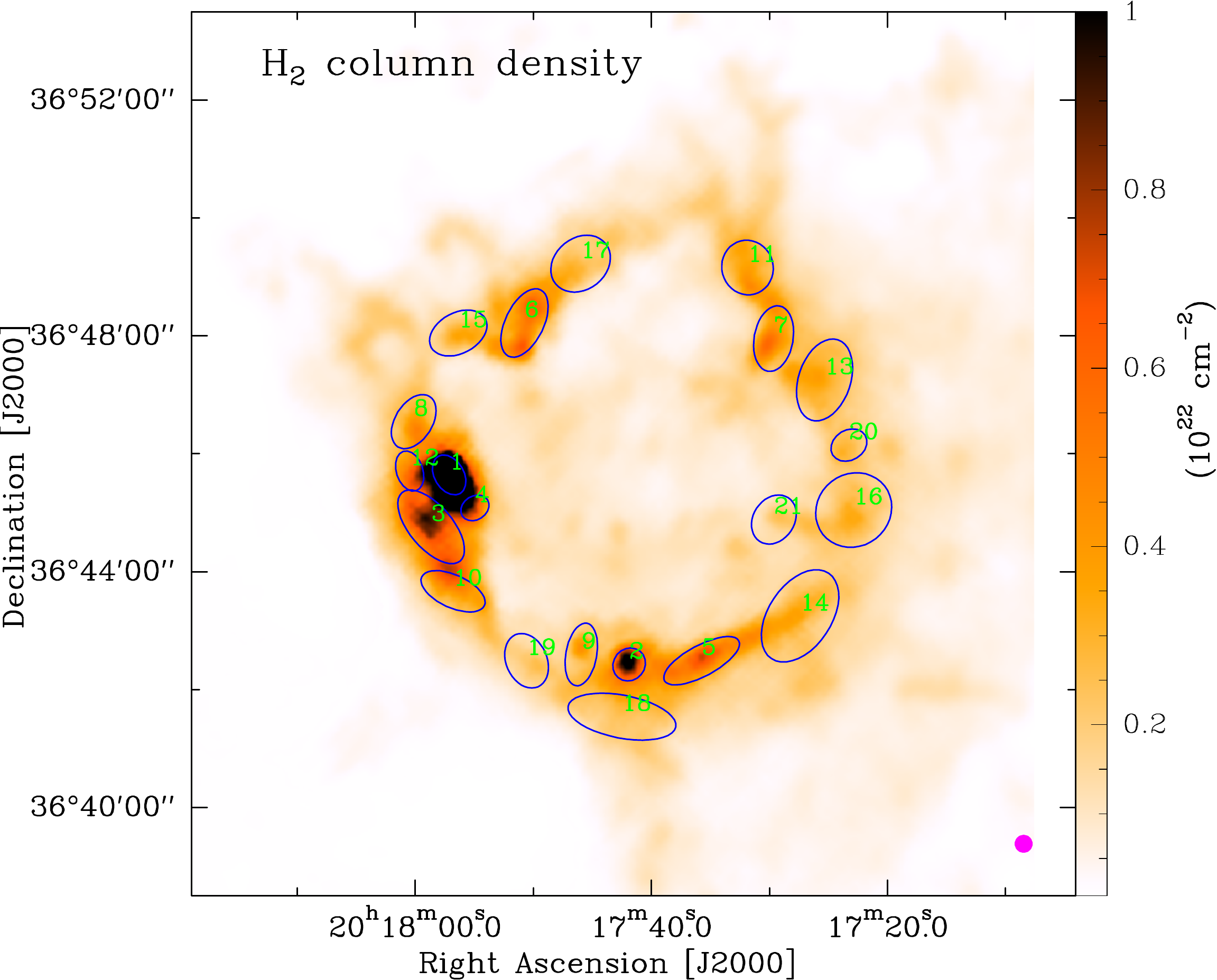}
\caption{High-resolution (18$\farcs2$) column density map of the Sh2-104 region derived from Hi-GAL survey data. The blue ellipses indicate the identified clumps. }
\label{Fig:dust-clumps}
\end{figure}

\begin{table}
\begin{center}
\tabcolsep 3mm\caption{Derived parameters of identified clumps}
\def\temptablewidth{8\textwidth}%
\vspace{-2mm}
\begin{tabular}{llcccccccccccr}
\hline\hline\noalign{\smallskip}
ID    & R.A. & Dec. & $\theta_{\rm min}$ & $\theta_{\rm maj}$ & $R_{\rm eff}$ & $N_{\rm H_{2}}$ & Mass & $n_{\rm_{H_{2}}}$ \\
   & (J2000)   & (J2000)   &  (arcsec)     &  (arcsec)  & (pc)  & (10$^{22}$ cm$^{-2}$)  & (M$_{\odot}$) 
& (cm$^{-3}$) \\
  \hline\noalign{\smallskip}       
  1  &20:17:54.7  &36:45:38.5  &30.0 &43.7  &0.30 &2.8 &251.1 & 5225 \\
  2  &20:17:42.5  &36:42:25.5  &31.8 &34.9	& 0.27	& 1.1	& 82.6	& 1912.2 \\ 
  3  &20:17:55.9  &36:44:45.7  &43.7 &91.3	& 0.58	& 1.0	& 330.0	& 1097.1 \\
  4  &20:17:53.0  &36:45:04.9  &24.1 &30.5	& 0.19	& 0.8	& 29.8	& 2317.7 \\
  5  &20:17:37.6  &36:42:29.3  &31.6 &85.9	& 0.45	& 0.7	& 151.4	& 1270.6  \\
  6  &20:17:49.6  &36:48:13.2  &39.1 &75.0	& 0.49	& 0.7	& 156.8	& 843.6  \\
  7  &20:17:32.7  &36:47:57.3  &39.8 &67.5	& 0.46	& 0.6	& 127.4	& 737.6  \\
  8  &20:17:57.1  &36:46:32.8  &37.9 &60.6	&  0.42	& 0.5	& 89.9	& 664.2 \\
  9  &20:17:45.7  &36:42:35.7  &31.6 &64.6	& 0.39	& 0.4	& 67.3	& 766.9 \\
 10  &20:17:54.4  &36:43:40.1  &52.2 &55.9	& 0.49	& 0.5	& 111.2	& 413.4  \\
 11  &20:17:34.5  &36:49:09.7  &27.0 &42.1	& 0.27	& 0.6	& 42.8	& 1328.1  \\
 12  &20:17:57.4  &36:45:42.4  &52.9 &86.2	&  0.63	& 0.4	& 155.9	& 353.1  \\
 13  &20:17:29.2  &36:47:15.2  &64.2 &104.6	&  0.77	& 0.4	& 218.1	& 263.1  \\
 14  &20:17:30.9  &36:43:15.0  &73.5 &79.8	& 0.72	& 0.3	& 169.3	& 202    \\
 15  &20:17:54.1  &36:48:03.0  &52.5 &65.4	& 0.54	& 0.3	& 88.7	& 273.9   \\
 16  &20:17:27.3  &36:45:02.7  &41.6 &57.8	& 0.44	& 0.3	& 51.1	& 312.3   \\
 17  &20:17:45.8  &36:49:13.5  &42.0 &53.1	& 0.42	& 0.2	& 41.5	& 273.4   \\
 18  &20:17:43.0  &36:41:32.4  &33.4 &70.1  &0.42	& 0.5	& 83.5	& 738.3  \\
 19  &20:17:49.5  &36:42:29.3  &41.5 &62.2  &  0.46	& 0.4	& 78.5	& 446.3   \\
 20  &20:17:27.6  &36:46: 8.9  &43.7 &111.7 & 0.64	& 0.3	& 121.3	& 328.5 \\
 21  &20:17:32.7  &36:44:53.1  &30.6 &38.7  &  0.28	& 0.2	& 19.2	& 439.3    \\     
\noalign{\smallskip}\hline
\end{tabular}\end{center}
\label{Table:dust-clump-1}
\end{table}

\begin{table}
\begin{center}
\tabcolsep 1.2mm\caption{Properties of identified clumps around the H {\small{II}} region Sh2-104}
\def\temptablewidth{8\textwidth}%
\vspace{-2mm}
\begin{tabular}{llcccccccccccr}
\hline\hline\noalign{\smallskip}
ID   &$T_{\rm mb}$(12)   &FWHM(12)  &$V_{\rm LSR}$(12) & $T_{\rm mb}$(13)   &FWHM(13)  &$V_{\rm LSR}$(13) & $T_{\rm ex}$& $\sigma_{\rm _{Therm}}$ & $\sigma_{\rm_{NT}}$ & $\sigma_{\rm_{v}}$ & $\alpha$\\
  & (K)   & (km s$^{-1}$)   &  (km s$^{-1}$)     &  (K)  & (km s$^{-1}$)  & (km s$^{-1}$) & (K) 
& (cm$^{-3}$) \\
  \hline\noalign{\smallskip}       
  1  &13.5(0.5)   &3.9(0.1)  &0.1(0.1) &5.4(0.2)    &2.6(0.1)   &-0.2(0.1)  & 17.0(0.5) & 0.22(0.01) & 1.10(0.02) & 1.12(0.01) & 1.8\\
  2  &12.0(1.2)   &3.5(0.2)  &2.5(0.1) &4.2(0.3)	&2.2(0.1)	& 2.5(0.1)	& 15.4(0.5)	& 0.21(0.01) & 0.92(0.03) &0.94(0.02) & 3.5\\ 
  3  &16.1(0.5)   &3.0(0.1)  &0.6(0.1) &6.1(0.2)	&1.9(0.1)	& 0.6(0.1)	& 19.6(0.5)	& 0.24(0.01) & 0.79(0.02) &0.83(0.01) & 1.5\\
  4  &17.2(0.5)   &3.3(0.1)  &0.4(0.1) &6.3(0.2)	&2.2(0.1)	& 0.2(0.1)	& 20.7(0.5)	& 0.25(0.01) & 0.94(0.02) &0.97(0.01) & 7.2\\
  5  &15.2(1.7)   &2.4(0.1)  &1.7(0.1) &4.6(0.3)	&1.8(0.1)	& 1.7(0.1)	& 18.7(0.5)	& 0.23(0.11) & 0.77(0.02) &0.80(0.01) & 2.3\\
  6  &15.2(1.2)   &3.3(0.1)  &0.6(0.1) &4.2(0.3)	&2.6(0.1)	& 0.5(0.1)	& 18.7(0.5)	& 0.23(0.01) & 1.09(0.03) &1.12(0.02) & 4.7\\
  7  &5.3(0.5)    &6.2(0.2)  &0.3(0.1) &2.1(0.3)	&2.4(0.2)	& 0.6(0.1)	& 8.6(0.5)	& 0.16(0.01) & 1.02(0.08) &1.03(0.04) & 4.6\\
  8  &9.7(0.5)    &4.0(0.1)  &0.1(0.1) &2.0(0.3)	&3.5(0.2)	& -0.1(0.1)	& 13.1(0.5)	& 0.20(0.01) & 1.47(0.07) &1.48(0.04) & 12.3\\
  9  &--    &-- &-- &--	&--	& --	&--	& -- &-- & -- &-- \\
 10  &9.1(0.8)    &2.8(0.2)  &1.0(0.1) &3.8(0.3)	&1.4(0.1)	& 1.1(0.1)	& 12.5(0.8)	& 0.19(0.01) & 0.59(0.03) &0.62(0.01) &2.0\\
 11  &9.9(0.5)    &3.1(0.1)  &1.0(0.1) &2.4(0.3)	&2.1(0.1)	& 0.8(0.1)	& 13.3(0.5)	& 0.20(0.01) & 0.90(0.05) &0.92(0.02) &6.4\\
 12  &8.5(0.5)    &4.4(0.1)  &-0.4(0.1) &2.8(0.3)	&3.2(0.1)	& -0.4(0.1)	& 11.8(0.5)	& 0.19(0.01) & 1.38(0.05) &1.40(0.02) &9.5\\
 13  &8.9(0.6)    &5.2(0.1)  &1.2(0.1) &2.3(0.3)	&2.4(0.2)	& 1.7(0.1)	& 12.3(0.6)	& 0.19(0.01) & 1.02(0.09) &1.04(0.04) &4.6\\
 14  &8.2(0.5)    &6.5(0.1)  &0.3(0.1) &2.3(0.6)    &1.7(0.2)	& 1.3(0.1)	& 11.6(0.5)	& 0.18(0.01) & 0.74(0.08) &  0.76(0.04) &2.9\\
 15  &15.6(0.5)   &3.0(0.1)  &0.4(0.1) &4.3(0.3)	&2.2(0.1)	& 0.3(0.1)	& 19.1(0.5)	& 0.24(0.01) & 0.92(0.02) &0.95(0.01) &6.6\\
 16  &3.6(0.5)    &9.7(0.1)  &0.7(0.1) &0.9(0.3)	&3.0(0.5)	& 3.7(0.2)	& 6.8(0.5)	& 0.14(0.01) & 1.27(0.20) &1.28(0.10) &16.9\\
 17  &8.9(0.6)    &3.3(0.2)  &0.1(0.1) &2.6(0.3)	&2.2(0.1)	& -0.2(0.1)	& 12.3(0.6)	& 0.19(0.01) & 0.91(0.05) &0.93(0.02) &10.5\\
 18  &16.8(1.2)   &2.7(0.1)  &2.4(0.1) &4.1(0.3)    &2.0(0.1)	& 2.6(0.1)	& 20.3(0.5)	& 0.24(0.01) & 0.82(0.03) &0.86(0.01) &4.5\\
 19  &6.0(0.5)    &2.8(0.1)  &0.7(0.1) &0.7(0.3)    &3.4(0.4)	& 0.5(0.2)	& 9.3(0.5)	& 0.17(0.01) & 1.43(0.18) &1.44(0.09) &14.6\\
 20  &4.1(0.5)    &7.3(0.2)   &-0.7(0.1)  &1.2(0.3) &2.2(0.3) & -2.2(0.1)	& 7.4(0.5)	& 0.15(0.01) &0.92(0.13)	& 0.93(0.07) & 5.5\\
 21  &--    &--  &-- &--   &--	& --	& --	& -- &-- &-- &--  \\     
\noalign{\smallskip}\hline
\end{tabular}\end{center}
\label{Table:dust-clump}
\end{table}

\subsection{Clumps Extraction}
Clumps with star formation activities are always found on the border of \HII region. The Hi-GAL survey data can be used to construct column density (N$_{H_{2}}$) map of the studied region. From the column density map, we can extract clumps. Previous method that the obtained column density maps is at the 36$\farcs3$ resolution of SPIRE 500 $\mu$m. To reveal more small structures, \citet{Palmeirimh2013} construct a column density map at the 18$\farcs2$ resolution of the SPIRE 250 $\mu$m data for the B211+L1495 region. Within the region covered by both PACS and SPIRE,  the reliability of the smoothed 250 $\mu$m version agrees with that  at the smoothed 500 $\mu$m.
Based on the smoothed 250 $\mu$m method of \citet{Palmeirimh2013}, the column density map of  Sh2-104  were created using a modified blackbody model to fit the SEDs pixel by pixel.  In Hi-GAL five bands, we only use  160, 250, 350 and 500 $\mu$m. Emission at 70 μm was excluded in the SED fitting because it can be contaminated by emission from small grains in hot PDRs. The dust opacity law  is assumed as $\kappa_{\upsilon}$ = 0.1$\times$(300 $\mu m /\lambda$)$^{\beta}$ cm$^{2}$/g, with $\beta$=2. The corresponding column density map of Sh2-104 is given in
Fig. \ref{Fig:dust-clumps}, which shows a ring-like structure. The column density of the ring ranges from $\sim 10^{21}$ cm$^{-2}$ to $\sim 10^{22}$ cm$^{-2}$. Using $^{13}$CO $J$=1-0 molecular line, the obtained mean column density of the ring is 6.8$\times10^{21}$ cm$^{-2}$, which is in agreement with that from Hi-GAL data.

 We use the algorithms Gaussclumps {\bf \citep{Stutzki1990,Kramer1998,Zhang2017}} to extract clumps and derive their physical properties from the column density map. Gaussclumps is a task included in the GILDAS package. Although Gaussclumps was originally written to decompose a three-dimensional data cube into Gaussian-shaped sources, it can 
also be applied to dust continuum or column density  maps without modification of the code. \citet{Kramer1998} suggested that “Stiffness” parameters that control the fitting were set to 1. A peak flux density threshold was set to 5$\sigma$. According to \citet{Belloche2011}, the initial guesses for the aperture cutoff, the aperture FWHM and the source FWHM were adopted as 8, 3, and 1.5 times the angular resolution, respectively. In Table \ref{Table:dust-clump-1}, we display all the clumps identified with Gaussclumps. Columns are (1) identification number of the clumps, (2)--(3) J2000 positions,  (4)–(5) the major and minor deconvolved FWHM of the clumps, (7) H$_{2}$ column density.

\subsection{Size, Mass and Volume Density}
Using  the major and minor deconvolved FWHM ($\theta_{\rm maj}$ and $\theta_{\rm min}$) of the clumps, the effective deconvolved radius of these clumps was determined by 
\begin{equation}
\mathit{R_{\rm eff}}=\sqrt{\theta_{\rm min}\times\theta_{\rm maj}}/2,
\end{equation}
The obtained $R_{\rm eff}$ are listed in column 6 of Table \ref{Table:dust-clump-1}. Clump radius are found to lie between 0.19 pc and 0.77 pc. 

In addition, the total mass of the clumps was calculated from their H$_{2}$ column density \citep{Kauffmann2008},
\begin{equation}
\mathit{M}=\mu m_{\rm H}\int N_{\rm H_{2}} \rm d \it A,
\end{equation}
Where $\mu$=2.72 is the mean molecular weight, $m_{\rm H}$ is the mass of  an H atom, and the surface element $\rm d \it A$ is ralated to the solid angle d$\Omega$ by $\rm d \it A$=$D^{2}$d$\Omega$, where $D$ is the distance of the clumps.

Assuming a 3D geometry for the clumps, the average volume density of each clump was calculated as 
\begin{equation}
\mathit{n_{\rm H_{2}}}=\frac{M}{4/3\pi R_{\rm min}^{2}R_{\rm maj}\mu m_{\rm H}},
\end{equation}
Where $R_{\rm min}$ and $R_{\rm maj}$ are  the respective minor and major axes of the clumps. The obtained volume densities are listed in Table \ref{Table:dust-clump-1}. 

\subsection{Velocity Dispersion and Virial Parameter($\alpha$)}
From H$_{2}$ column density map, 21 clumps are identified. Based on their coordinates and size, we search for $^{12}$CO $J$=1-0 and $^{13}$CO $J$=1-0 spectra that are located at or near the peak position of the clumps. Morever, we fitted the spectrum of each clump  with the Gaussian profile. The fitted parameters are listed in Table \ref{Table:dust-clump}, including brightness temperature ($T_{\rm mb}$), FWHM, and centroid velocity ($V_{\rm LSR}$). For clumps 9 and 21, we did not obtain the effective spectral value because of the weak signal ($\leq 3 \sigma$). Since the $^{13}$CO $J$=1-0 emission is optically thin in the ring (see Sec. 3.3), we use the FWHM of $^{13}$CO $J$=1-0 to calculate the one-dimensional velocity dispersion ($\sigma_{\upsilon}$). Because these clumps are distributed mainly over the molecular ring associated with Sh2-104, we also need consider the thermal velocity dispersion. $\sigma_{\upsilon}$  in each clump is calculated as follows: 
\begin{equation}
\mathit{\sigma_{\upsilon}}=\sqrt{\sigma_{\rm Therm}^{2}+\sigma_{\rm NT}^{2}}
\end{equation}
Where $\sigma_{\rm Therm}$ and $\sigma_{\rm NT}$ are the thernal and non-thermal one-dimensional velocity dispersions. For all the clumps, $\sigma_{\rm Therm}$ and $\sigma_{\rm NT}$ can be derived as, respectively:
\begin{equation}
\mathit{\sigma_{\rm Therm}}=\sqrt{\frac{kT_{\rm ex}}{m_{\rm H}\mu}}
\end{equation}
\begin{equation}
\mathit{\sigma_{\rm NT}}=[\sigma_{\rm ^{13}CO}^{2}-\frac{kT_{\rm ex}}{m_{\rm ^{13}CO}\mu}]^{1/2}
\end{equation}
Where $T_{\rm ex}$ is the excitation temperature of the clumps. We used $^{12}$CO $J$=1-0 to calculate $T_{\rm ex}$ following equation 2.  $\sigma_{\rm ^{13}CO}$ = ($\Delta V_{13}/\sqrt{\rm 8ln2}$) is the one-dimensional velocity dispersion of $^{13}$CO $J$=1-0 while $m_{\rm ^{13}CO}$ is the mass of $^{13}$CO $J$=1-0. The derived parameters are summaried in Table \ref{Table:dust-clump}. 

Stability of the clumps against gravitational collapse can be evaluated using the virial parameter \citep{Kauffmann2013}
\begin{equation}
\mathit{\alpha}=1.2(\frac{\sigma_{\upsilon}}{\rm km \ s^{-1}})^{2}(\frac{R_{\rm eff}}{\rm pc})(\frac{M}{10^{3}M_{\odot}})^{-1}
\end{equation}
Where $\sigma_{\upsilon}$ is the one-dimensional velocity dispersion, $R_{\rm eff}$ is the effective deconvolved radius, and $M$ is the mass of the clumps.  If $\alpha \leq 2$, the clumps collapse \citep{Bertoldi1992,Kauffmann2013}. Hence, from Table \ref{Table:dust-clump}, we obtain that clumps 1, 3, and 10 collapse. 

\begin{figure}
\centering
\includegraphics[width = 0.45 \textwidth]{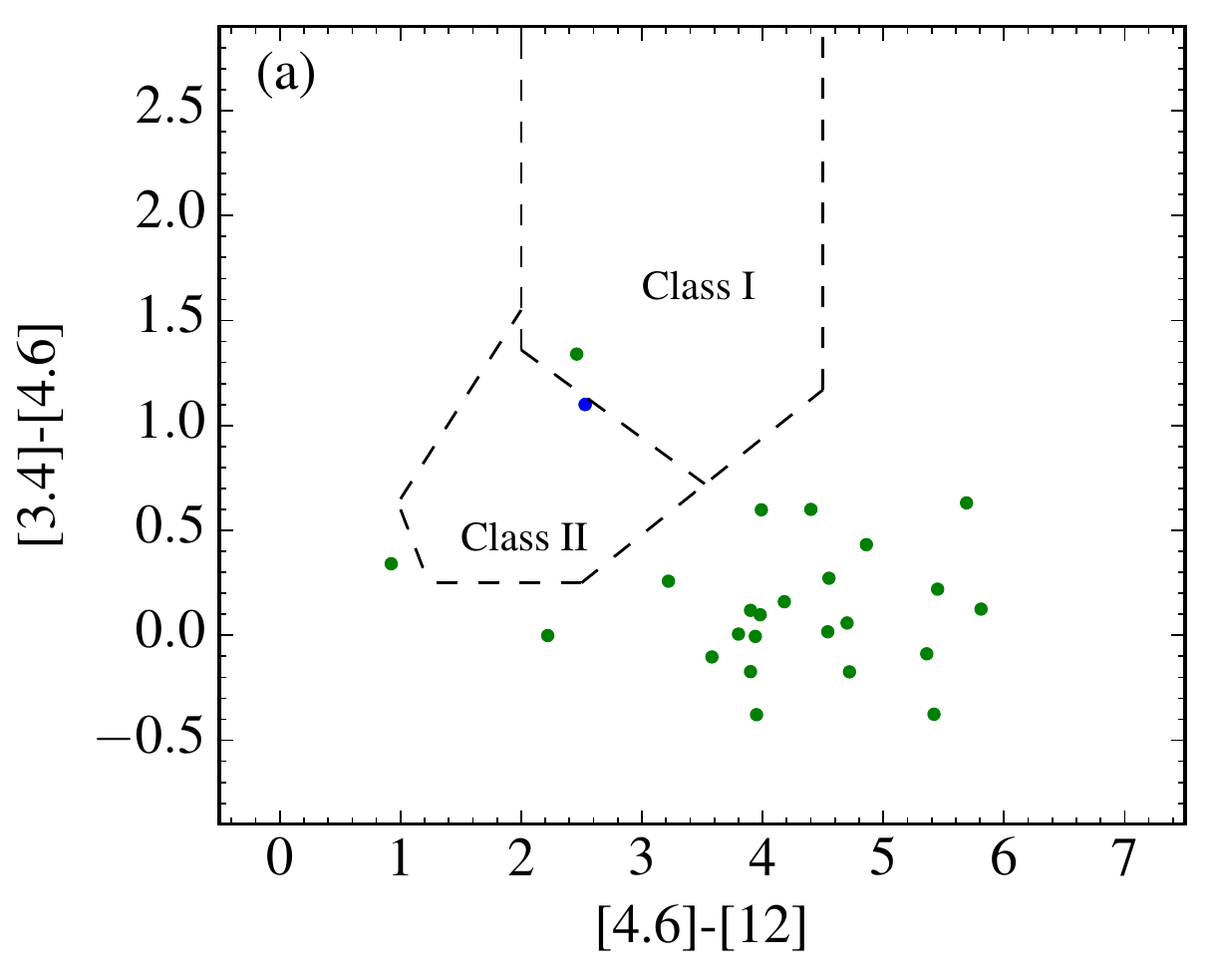}
\includegraphics[width = 0.45 \textwidth]{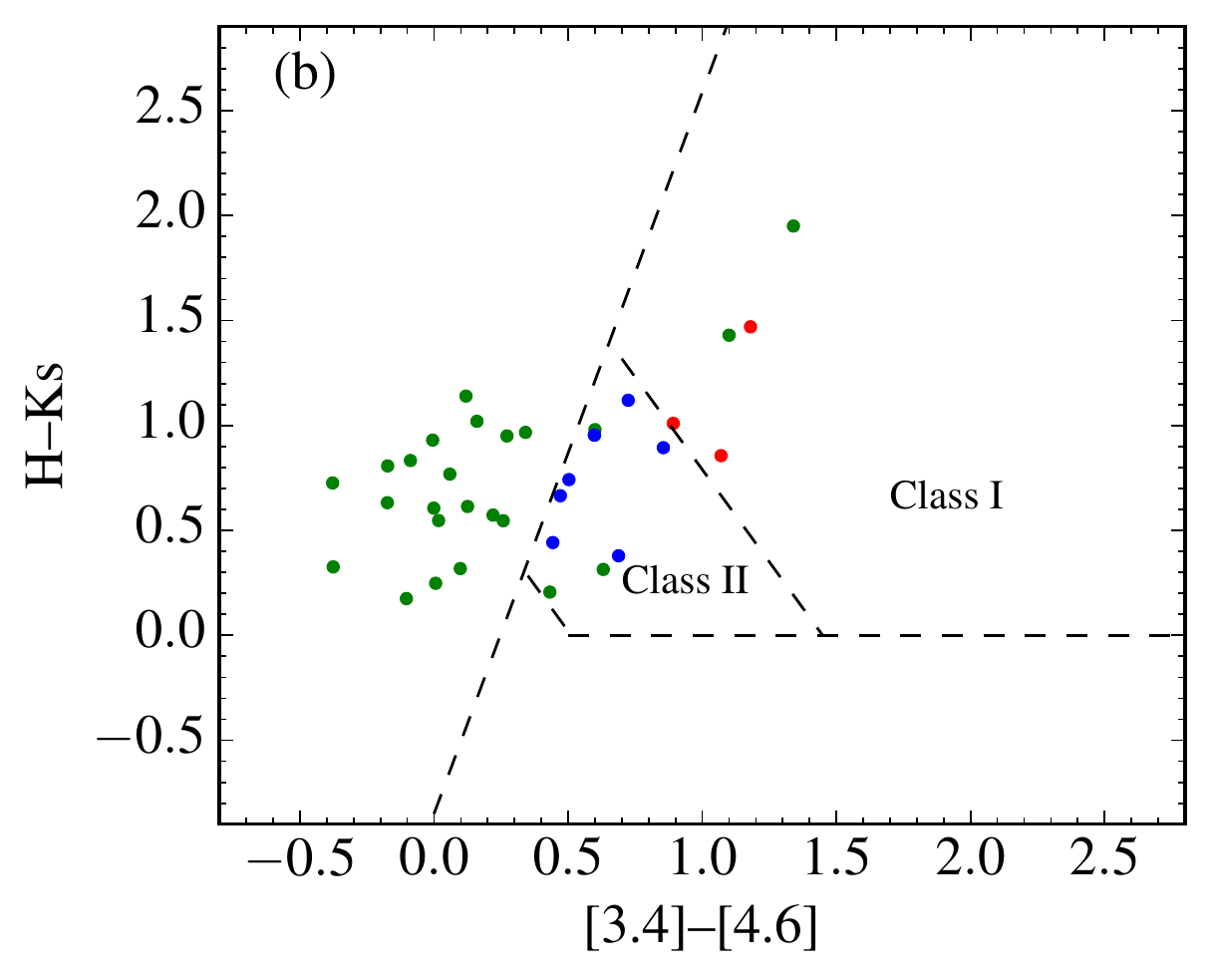}
\caption{Left panel: WISE bands 1, 2, 3 color-color diagram. Right panel: WISE bands 1 and 2 and 2MASS H and Ks color–color diagram. Dashed lines indicate the boundaries by which we classify Class I and Class II. Class I YSOs are labelled as red dots, and class
II YSOs as blue dots. The green dots indicate Class I/II YSOs from \citep{Marton2016}. }
\label{Fig:color}
\end{figure}

\begin{figure}
\centering
\includegraphics[width = 0.5 \textwidth]{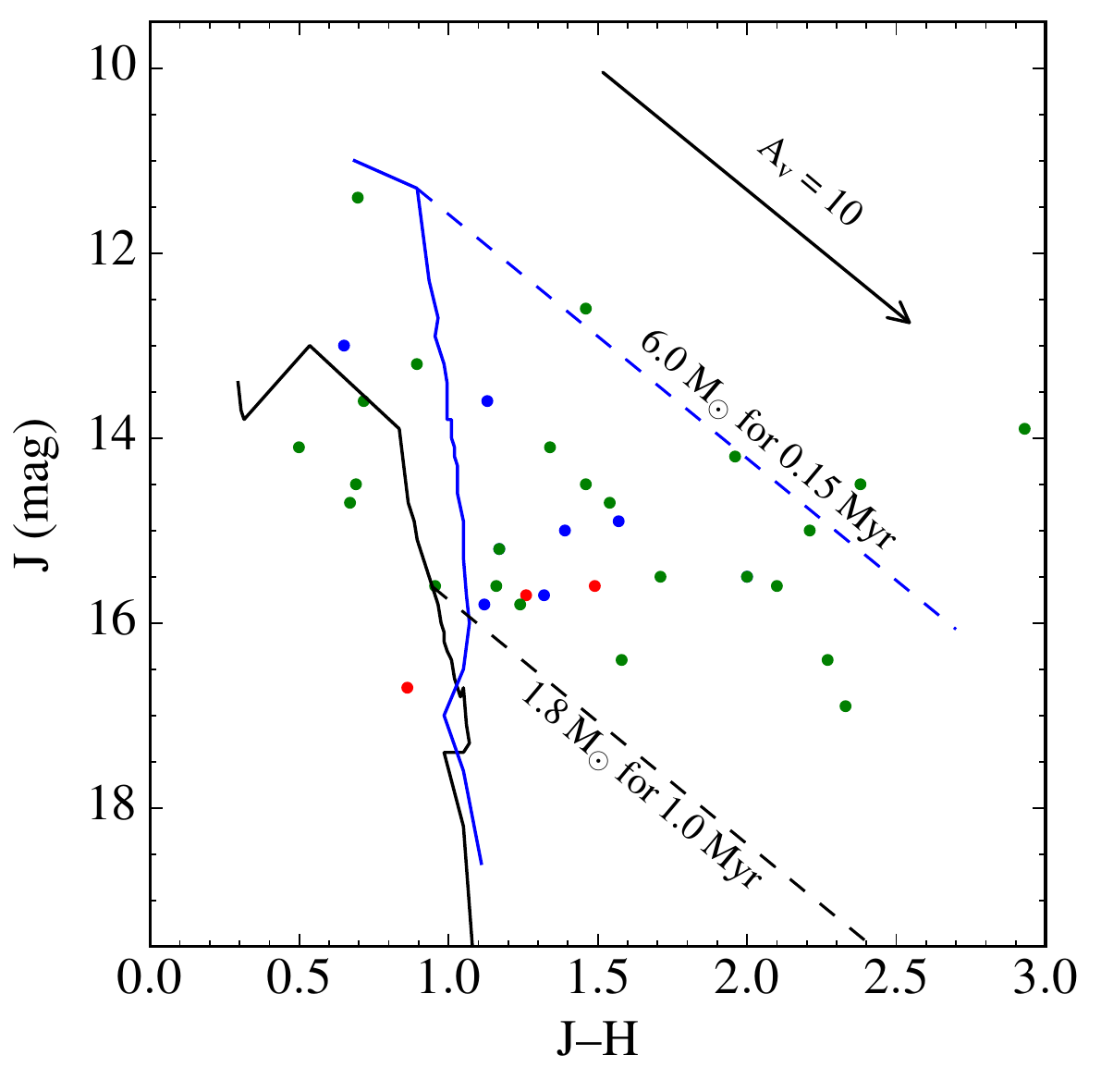}
\vspace{-4mm}
\caption{J/J-H colour-magnitude diagram for the YSOs.  Class II YSOs are labelled as red dots, and Class I YSOs as  blue  dots. The green dots indicate Class I/II YSOs from \citep{Marton2016}. The black solid curve denotes the location of 10$^{6}$ yr old pre–main-sequence (PMS) stars, and the blue solid curve is for those that is 1.5$\times$10$^{6}$ yr old, both were obtained from the model of \citet{Siess2000}.  All the isochrones are corrected for the distance and reddening of the Sh2-104 region. Dashed lines represent reddening vectors for various masses. The black dashed line denotes the position of a PMS star of 1.8 M$_{\odot}$ for 1.0 Myr, and the blue dashed lines indicate the position of star of 6.0 M$_{\odot}$ for 0.15 Myr.}
\label{Fig:JH-J}
\end{figure}

\begin{figure}
\centering
\includegraphics[width = 0.5 \textwidth]{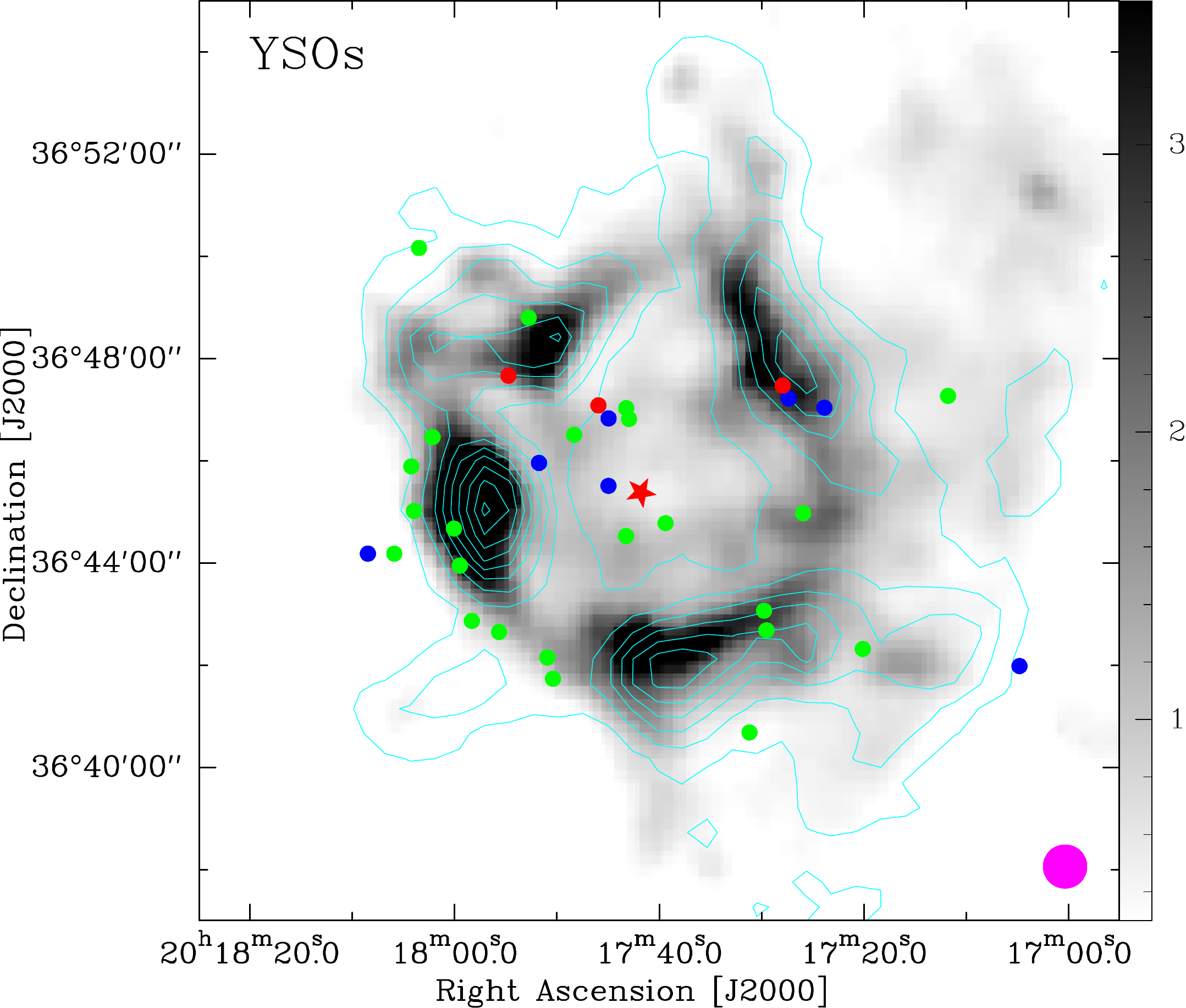}
\caption{$^{13}$CO $J$=1-0 integrated intensity map overlaid on the  Hersel 500 $\mu$m image (grey scale). Class II YSOs are labelled as red dots, and Class I YSOs as  blue  dots. The green dots indicate Class I/II YSOs from \citep{Marton2016}. The unit for the colour bar is Jy pixel$^{-1}$.}
\label{Fig:YSO}
\end{figure}

\subsection{Distribution of Young Stellar Objects}
To detect young stellar objects (YSOs) adjacent to Sh2-104, we use the WISE and 2MASS all-sky source  catalogs \citep{Skrutskie2006,Wright2010}.  From these two catalogs, we select the sources with the 3.4, 4.6, 12, and 22 $\mu$m, and $J$, $H$, and $Ks$ bands within a circle of 9$^{\prime}$ in radius centered on the ionized star of Sh2-104. We select YSOs using the criteria described in \citet{Koenig2014}, as shown in Fig. \ref{Fig:color}. This criteria is based on the $[4.6]-[12.0]$ versus $[3.4]-[4.6]$ color-color, and $[Ks]-[3.4]$ versus $[3.4]-[4.6]$ color-color diagrams. For many objects visible in WISE bands 1 and 2 will lack a reliable band 3 or 4 detection,  the $[3.4]-[4.6]$ versus $[H]-[Ks]$ color-color diagram is a supplement to $[4.6]-[12.0]$ versus $[3.4]-[4.6]$ color-color diagram. Based on above color selection criteria of YSOs, we only find 3 Class I YSOs and 8 Class II YSOs. Furthermore, a much cleaner selection technique has been used by \citet{Marton2016} to create an all-sky catalogue of YSOs (Class I/II). According to the catalogue of  \citet{Marton2016}, we find 23 Class I/II YSOs. Comparing the YSO positions, only an YSO overlap between the two methods. Class I YSOs are protostars with circumstellar envelopes, while Class II YSOs are disk-dominated objects \citep{Evans2009}. The $J$ luminosity is used to  estimate stellar masses, as $J$ is less affected by the emission from circumstellar material \citep{Bertout1988, Yadav2016}. Figure \ref{Fig:JH-J} shows the $J/J-H$ colour-magnitude diagram for the selected YSOs. The black solid curve denotes the location of 10$^{6}$ yr old pre-main-sequence (PMS) stars, and the blue solid curve is for those that is 1.5$\times$10$^{6}$ yr old, both were obtained from the model of \citet{Siess2000}.  All the isochrones are corrected for the distance and reddening of the Sh2-104 region. Dashed lines represent reddening vectors for various masses. The black dashed line denotes the position of a PMS star of 1.8 M$_{\odot}$ for 1.0 Myr, and the blue dashed lines indicate the position of star of 6.0 M$_{\odot}$ for 0.15 Myr. From Fig. \ref{Fig:JH-J}, we see that the majority of the YSOs have masses in the range 1.8-6.0 M$_{\odot}$, and an age range of 0.15-1.0 Myr. Figure \ref{Fig:YSO} shows the spatial distribution of both Class I and Class II YSOs.  From Fig. \ref{Fig:YSO}, we also note that  Class I/II YSOs are mostly concentrated along the molecular and dust ring.

\begin{figure*}
\centering
\includegraphics[width = 1.0 \textwidth]{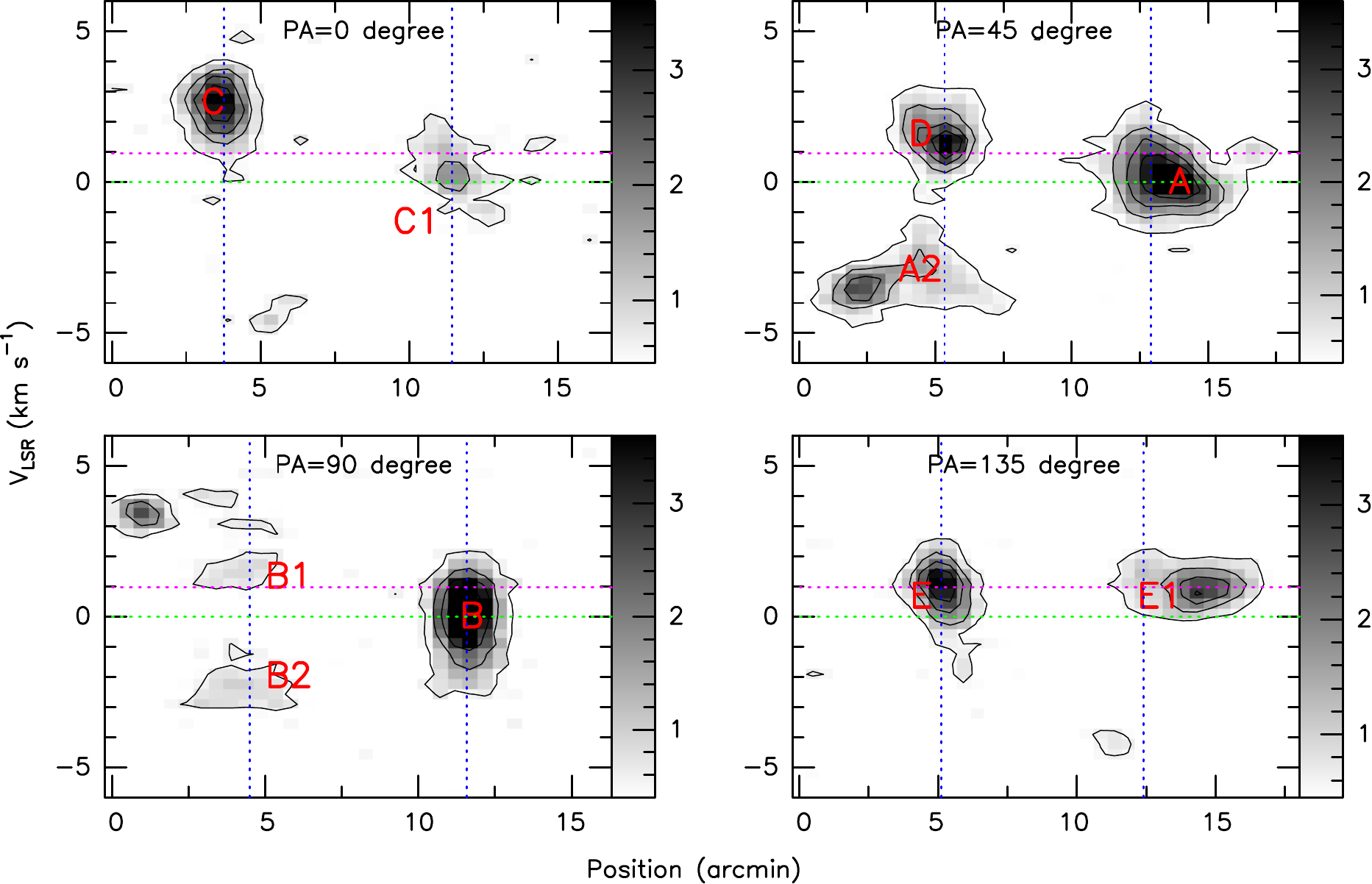}
\caption{P-V diagrams constructed from  $^{13}$CO $J=1-0$ transition for different directions. The contour levels are 10, 20,..., $90\%$ of the peak value. The blue dashed lines show the projected position of each edge in the different position angle, while the green dashed lines indicate the systemic velocity (0 km s$^{-1}$) of Sh2-104; The pink dashed lines indicate the systemic velocity of 1 km s$^{-1}$.}
\label{Fig:13CO-PV}
\end{figure*}

\section{Discussion}
\label{sect:discu}
\subsection{Gas Structure Around Sh2-104 }
The ionized gas of Sh2-104 shows a  ring-like structure with a radius of about 2.9 pc (2.5\amin at 4 kpc), which is consistent with the optical image \citep{deha2003}. \citet{deha2010} studied 102 bubbles.  Only bubbles N49 and N61 show a shell-like structure in both the radio continuum emission and the 24 $\mu$m emission \citep{Watson08,Xu2016}. They suggested that such shell-like structure could be due to a stellar wind emitted by the central massive star. Both CO molecular gas and cool dust emission shows a ring-like shape, which just encloses the ionized gas of Sh2-104. If an \HII region spherically expands,  both the collected gas and dust emission will also show three-dimensional spherical shell structure. However, \citet{beau2010} suggested that the collected gas and dust emission around \HII region is two-dimensional rings with thicknesses not greater than the ring sizes.  For the three-dimensional shells, expanding along the line of sight will result in blue-shifted emission from the near-side of the shell, and red-shifted emission from the far side \citep{Anderson2015}. Here for the CO molecular ring surrounded \HII region Sh2-104, we also did not detect the CO emission of higher than 3 signal (0.6 K) inside the ring. 

\begin{figure*}
\centering
\includegraphics[width = 0.48 \textwidth]{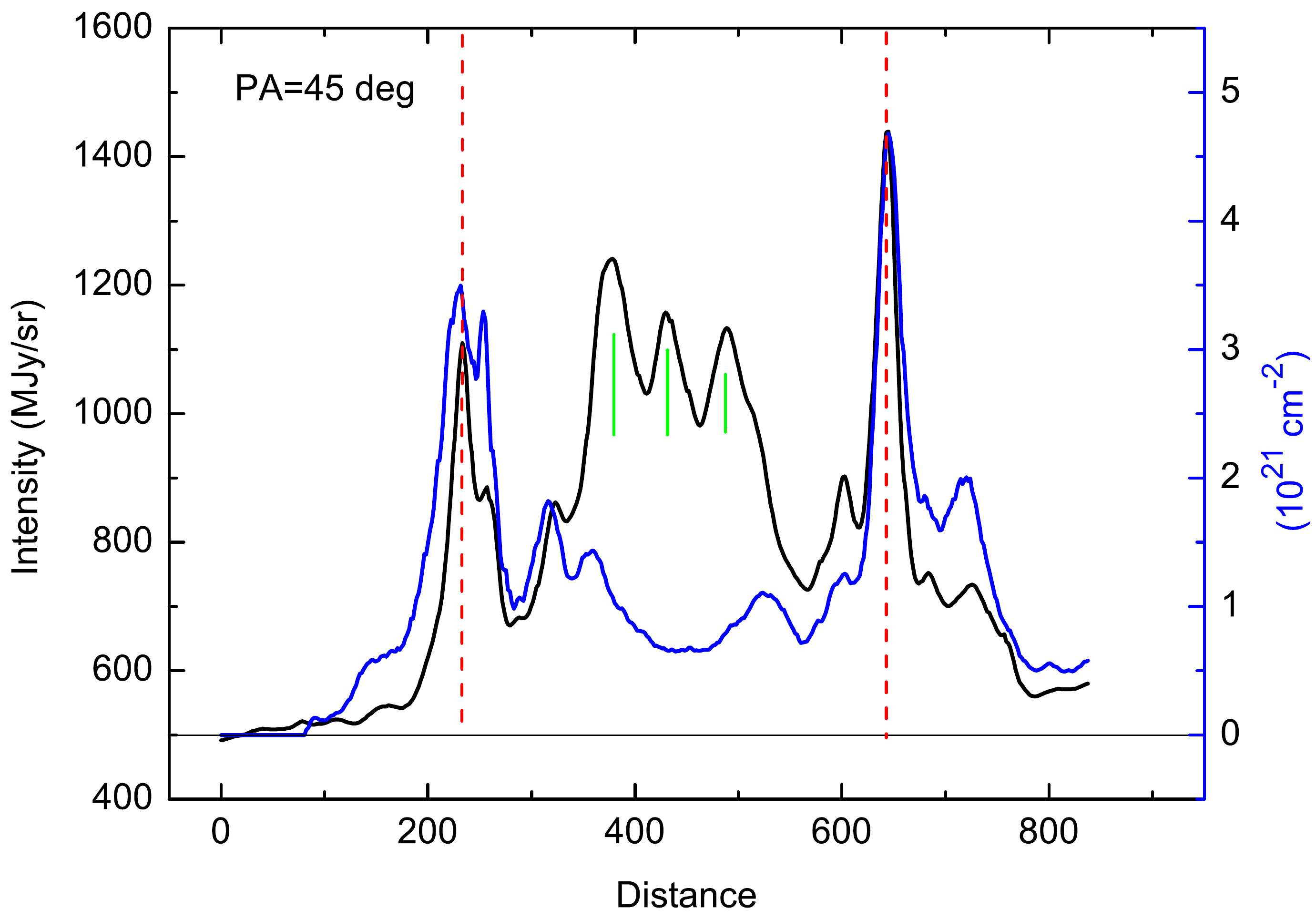}
\includegraphics[width = 0.48 \textwidth]{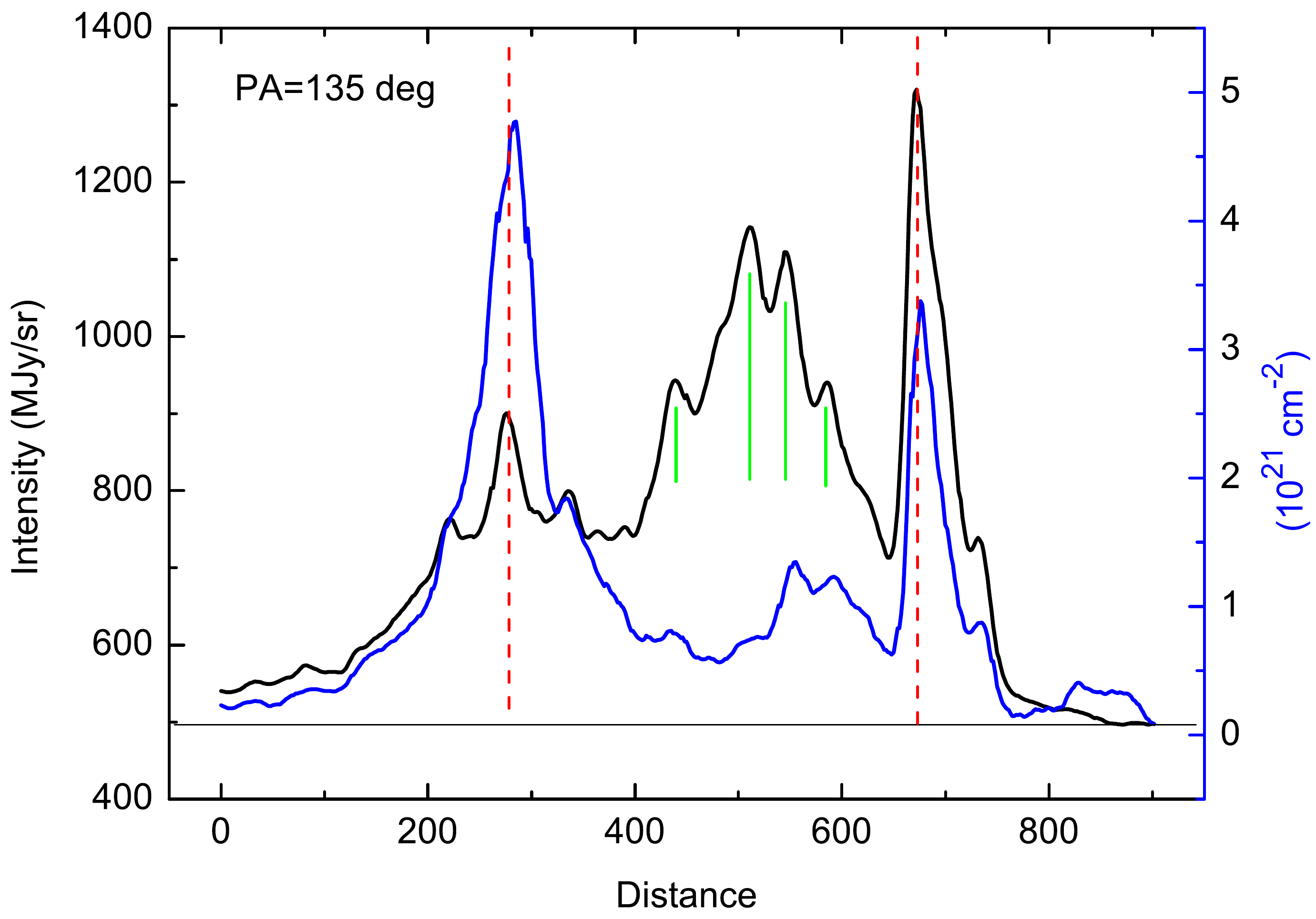}
\caption{Slice profiles of WISE 12 $\mu$m emission (black lines) and H$_{2}$ column density map (blue lines)  through Sh2-104 in PA = 45degree and PA = 135 degree directions. The cutting paths are shown as the green arrows in Fig. \ref{Fig:13CO-PV}. The red dashed lines mark the two edges of Sh2-104, while the green lines may indicate some gas emission. }
\label{Fig:S104-slice}
\end{figure*}

To further reveal the structure of the ring, we made four position-velocity (PV) diagrams in $^{13}$CO $J$=1-0 line, as shown in  Fig. \ref{Fig:13CO-PV}. These PV diagrams are obtained from the four cuts through the position of the exciting star with the position angles (PA) of 0$\adeg$, 45$\adeg$, 90$\adeg$, and 135$\adeg$ (see Fig. \ref{Fig:S104-13CO-500}). From Fig. \ref{Fig:13CO-PV}, we see that the symmetrical borders of the ring are detected along all the four cuts.  Each border is marked by a blue dashed line. Between two symmetrical borders, we still did not detect the CO emission. Since our selected cuts just go through these identified clumps, then the position of each clump can indicate a border of the ring. For the symmetrical borders, we design name at clumps A1, B1, C1, D1, and E1. If one border has two  CO components, the other component will be named such as clumps A2 and B2.  Through inspected the clumps A2 and B2 with association of the Sh2-104 morphology, we suggest that there two clumps  should belong to the foreground emission of the corresponding border, and only overlap with the other component in the line of sight.  Through the PV diagrams (Fig. \ref{Fig:13CO-PV}), we found that both the borders of the ring with clumps E and E1 have a systemic velocity of about 1 $\kms$. The cut line through clumps E and E1 divide the ring into the north-east and south-west two portions. The north-east portion containing the clumps A, B, and C1 is blueshifted, while the south-west portion  with the clumps B1, C, and D is redshifted. Hence, we concluded that the ring is expanding with an inclination relative to the  plane of sky. The existence of the inclination indicates that the ring is not spherically symmetrical.  Additionally, from Fig. \ref{Fig:13CO-PV}, we also see  that the projected distances are the same and 7.5$\amin$ between two symmetrical edges along the different PAs. The distance between clumps E and E1 is roughly equal to  their projected distance. While the distances between clumps C and C1, clumps A and D, clumps B and B1 are greater than 7.5$\amin$, because the cut lines through two symmetrical clumps have an inclination angle with the projected axis. It further suggests that the ring is not spherically symmetrical. The ionized gas of Sh2-104 has a LSR velocity of about 0 $\kms$ \citep{Georgelin1973}. While the expansion of the molecular ring begin at the LSR velocity of 1 $\kms$. If there is a three-dimensional spherical shell that just encloses Sh2-104, the ionized gas and the ring should have the same LSR velocity.  From above analysis,  we concluded that the collected gas emission around \HII region Sh2-104 may be a two-dimensional ring-like structure. 

In addition, we also make the slice profiles of WISE 12 $\mu$m emission (black lines) and H$_{2}$ column density map (blue lines) through Sh2-104 in PA = 45 degree and PA = 135 degree directions, as shown in Fig. \ref{Fig:S104-slice}, which shows the 12 $\mu$m flux and H$_{2}$ column density change as a function of the position. The two symmetrical borders (red dashed lines) of the ring related to Sh2-104 display the strang 12 $\mu$m and H$_{2}$  emission, indicating that the strang PAHs molecules traced by the 12 $\mu$m emission are excited only if there should be some molecular gas adjacent to H II region. Howerver, the 12 $\mu$m emission is very strang inside the ring, while the H$_{2}$ column density is lower in the corresponding positions. Compared with the 70 $\mu$m emission in the left panel of Fig. \ref{Fig:S104-70}, we that found the 12 $\mu$m emission is associated with that  at 70 $\mu$m inside the ring, suggesting that this part of the 12 $\mu$m emission is also producted from the hot dust. Thus, it further confirms that the collected gas emission is likely to  a two-dimensional ring-like structure around \HII region Sh2-104.

\subsection{Star Formation Scenario}
\begin{figure}
\centering
\includegraphics[width = 0.6 \textwidth]{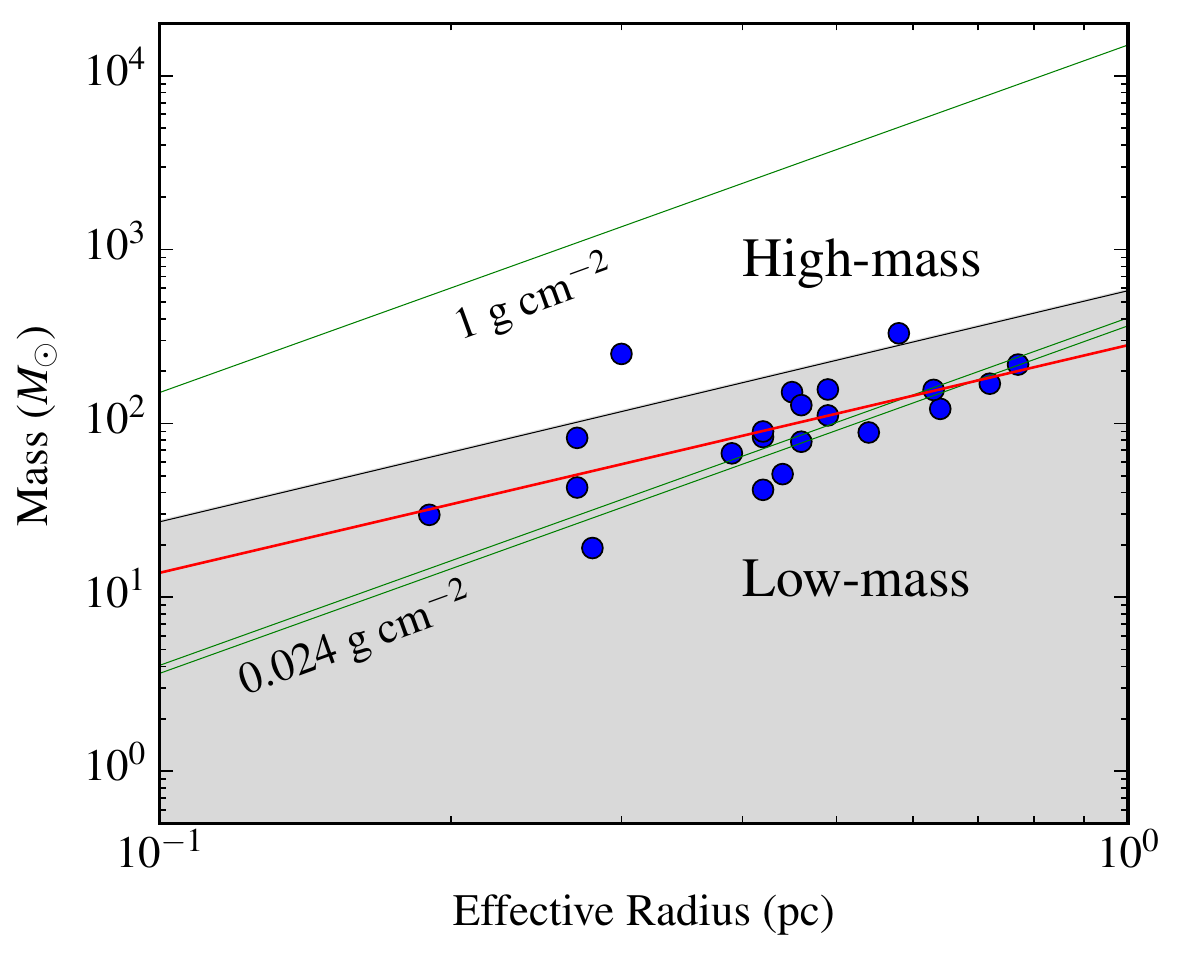}
\vspace{-4mm}
\caption{Mass versus radius relationship. The grey region represents the parameter space to be devoid of massive star formation. The black line delineates the threshold (Kauffmann \& Pillai 2010). The blue solid circles indicate the clumps. Upper and lower solid green lines present surface densities of 1 g cm$^{-2}$ and 0.024 g cm$^{-2}$, respectively. The red line shows the result of a power-law fit to all the clumps, where we find $M(r)=281M_{\odot}(r/\rm pc)^{1.31\pm0.08}$.}
\label{Fig:mass-size}
\end{figure}

Five large clumps are regularly spaced along the ring.  One of these clumps is associated with an $\uchii$ region and IRAS 20160+3636. \citet{deha2003} suggested that the CC process is at work in this region. From the high-resolution (18$\farcs2$) column density map constructed by the Hi-GAL survey data, we extract 21 clumps. To determine whether the identified clumps have sufficient mass to form massive stars, we must consider their sizes and mass. According to \citet{Kauffmann2010}, if the clump mass is $M(r)\geq580M_{\odot}(r/\rm pc)^{1.33}$, {\bf thus} they can potentially form massive stars. Figure \ref{Fig:mass-size} presents a mass versus radius plot of the clumps. We find that only two clumps (\#1 and \#3) above the threshold, indicating that these two clumps are dense and massive enough to potentially form massive stars. About 90\% of all the identified clumps will form low-mass stars. Moreover, we evaluate the  stability of these clumps against gravitational collapse using the virial parameter, finding only three clumps collapse, including clumps \#1 and \#3. The surface density of  0.024 g cm$^{-2}$ gives the average surface density thresholds for efficient star formation. The thresholds, shown in green lines in Fig. \ref{Fig:mass-size},  are derived by \citep{Lada2010,Heiderman2010}, respectively. As can be seen from Fig. \ref{Fig:mass-size}, 15 clumps (72\%) lie at or above the lower surface density limit of  0.024 g cm$^{-2}$. A power-law fit to the identified clumps yields $M(r)=281M_{\odot}(r/\rm pc)^{1.31\pm0.08}$. The slope we obtained is equal to what was found in \citet{Kauffmann2010}.

The selected YSOs (Class I and Class II ) are mostly concentrated in the whole ring around \HII region Sh2-104. The high density of YSOs located in the ring that these YSOs are physically associated with Sh2-104.  The dynamical age of  Sh2-104 can be used to  decide whether these YSOs are triggered by the \HII region. Using a simple model described by \citep{Dyson80} and assuming an H {\small II} region expanding in a homogeneous medium, we estimate the dynamical age of the \HII region as
\begin{equation}
\mathit{t_{\rm  dyn}}=\frac{4R_{\rm s}}{7c_{\rm s}}[(\frac{R}{R_{\rm s}})^{7/4}-1],
\end{equation}\indent
where $R_{\rm s}$ is the radius of the Str\"omgren sphere given by $R_{\rm s}$ = $3Q_{\rm Ly}/4\pi n_{0}^{2}\alpha_{\rm B}$, where $Q_{\rm Ly}$ is the ionizing luminosity,  $n_{\rm 0}$ is the initial number density of the ambient medium around the \HII region, $\alpha_{\rm B}$ = 2.6$\times$10$^{-13}$ cm$^{3}$ s$^{-1}$ is the hydrogen recombination coefficient to all levels above the ground level. For the ionized medium we adopt a  sound velocity of $c_{\rm s}$ $\simeq$ 10 km s$^{-1}$. Moreover, we adopt the radius (3.7$\amin$) of the ring  as that of \HII region Sh2-104, which is obtained from Fig. \ref{Fig:13CO-PV}. Taking the distance of $\sim$4 kpc to Sh2-104  \citep{deha2003}, the  \HII region radius is 4.4 pc.  As  a rough estimate, $n_{\rm 0}$ can be determined  by  distributing the total molecular mass over a sphere \citep[e.g.,][] {Zavagno2007,Paron2009,Cichowolski2009,Anderson2015,Duronea2015}. The mass of the ring is $\sim$2.2$\times10^{4}\rm M_{ \odot}$. Using this mass, we  only consider the number of atoms that can be ionized, so this would imply that  $n_{\rm 0}$ = 2.6$\times10^{3}$ \cmcub. \HII region Sh2-104 is excited by an O6V center star \citep{Crampton1978,Lahulla1985}. For an O6V star, the ionizing luminosity ($Q_{\rm Ly}$) is 10$^{48.99}$ $\rm s^{-1}$ \citep{Martins2005}. \citet{Inoue2001} suggested that only half of Lyman continuum photons from  the central source in a Galactic  \HII region ionizes neutral hydrogen, the remainder being absorbed by dust grains within the ionized region. Finally,  we obtain the ionizing luminosity of 10$^{48.69}$ $\rm s^{-1}$. Hence, we derived a dynamical age of 1.6$\times10^{6}$ yr for \HII region Sh2-104. Additionaly, from the $J/J-H$ colour-magnitude diagram for the selected YSOs, we derive that the majority of the YSOs have masses in the range 1.8-6.0 M$_{\odot}$, and an age range of 0.15-1.0 Myr. Comparing the dynamical age of \HII region Sh2-104 with that of YSOs shown on the ring, we conclude that these YSOs are likely to be triggered by \HII region Sh2-104. It also means that the CC or RDI processes induce the formation of these YSOs. \citet{deha2003} suggested     \HII region Sh2-104  is a typical candidate for triggering star formation by CC process.

\begin{figure}
\centering
\includegraphics[width = 0.6 \textwidth]{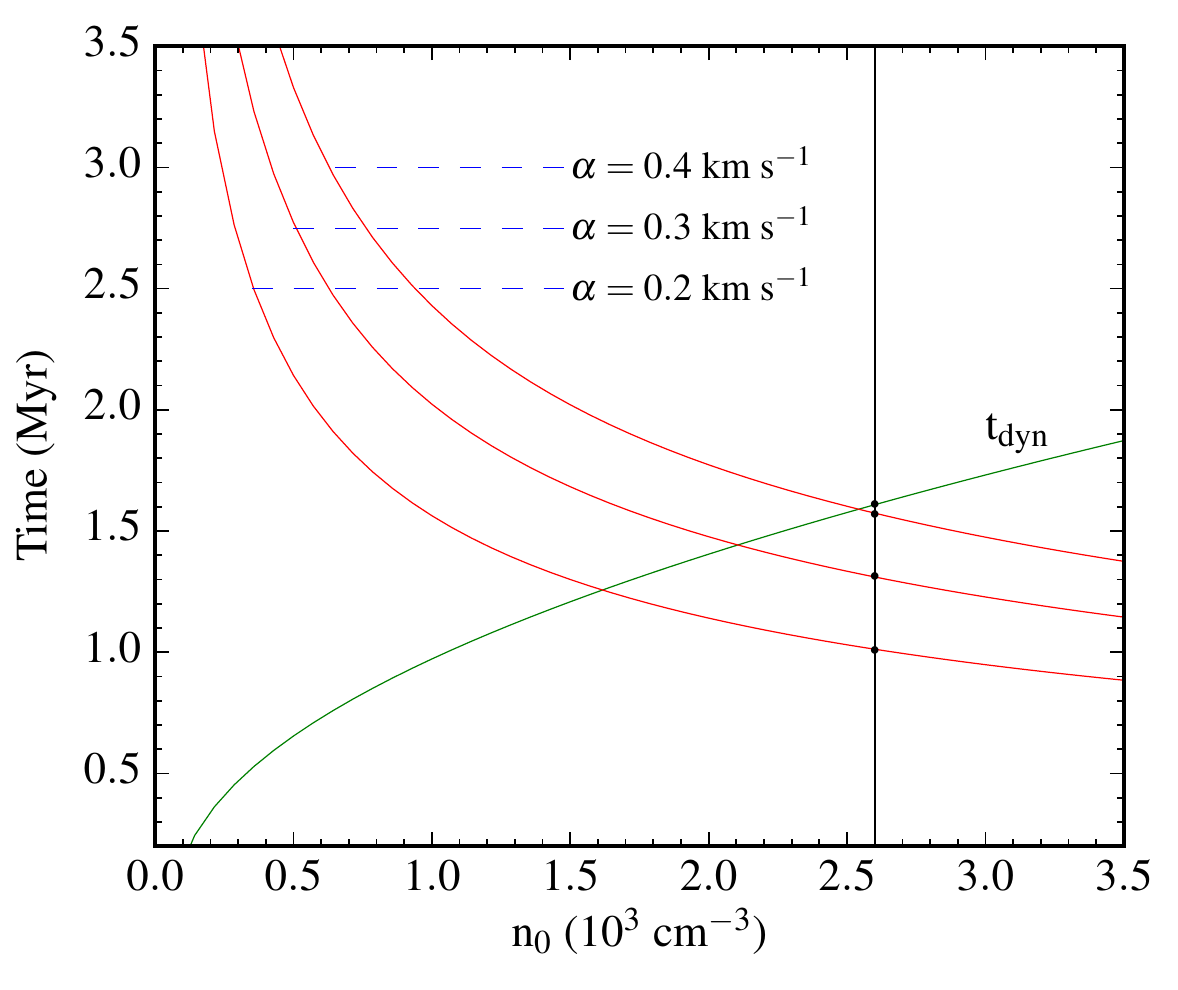}
\vspace{-4mm}
\caption{plot of dynamical time (green curve) and fragmentation timescale (red curves) as a function of initial density of the ambient medium. The fragmentation timescale is calculated for different sound speeds of neutral gas ($\alpha_{s}$ = 0.2, 0.3, and 0.4 km s$^{-1}$).  The black vertical line marks the position of n$_{0}$ = 2.6$\times10^{3}$ cm$^{-3}$.}
\label{Fig:THII-age}
\end{figure}

The expansion of Sh2-104 have collected the gas into the ring. To determine whether the fragmentation occurs around Sh2-104 via the CC process, we compute the fragmentation time of the collected gas based on the theoretical model of  \citet{Whitworth1994} :
\begin{equation}
\mathit{t_{\rm  frag}}=1.56(\frac{\alpha_{\rm s}}{0.2})^{7/11}(\frac{Q_{\rm Ly}}{10^{49}})^{-1/11}(\frac{n_{0}}{10^{3}})^{-5/11} \rm \ Myr,
\end{equation}\indent
The diagram of $t_{\rm frag}$ and $t_{\rm dyn}$  as a function of the initial ambient density $n_{0}$ is shown in Fig. \ref{Fig:THII-age}. In comparison, we find that the dynamical age of \HII region Sh2-104 ($t_{\rm dyn}$ $\approx$ 1.6 Myr) is greater than the fragmentation time.  This indicates that the ring of the collected gas has had enough time to fragment into the identified clumps during the lifetime of \HII region Sh2-104. Thus, it further confirm that the CC process is presumably at work \HII region Sh2-104.

\section{CONCLUSIONS}
\label{sect:summary}
We present the molecular $^{12}$CO $J$=1-0, $^{13}$CO $J$=1-0 and C$^{18}$O $J$=1-0, infrared, and  radio continuum observations toward \HII region Sh2-104. The main results are summarized as follows:

1.  \HII region Sh2-104 shows a double-ring structure, the outer ring traced by 12 $\mu$m, 500 $\mu$m, $^{12}$CO $J$=1-0, and $^{13}$CO $J$=1-0 emission, and the inner ring traced by by 22 $\mu$m and 21 cm emission. We suggest that the outer ring with a radius of 4.4 pc is created by the expansion of \HII region Sh2-104, while the inner ring with a radius of 2.9 pc may be produced by the energetic stellar wind from its central ionized star. We derived a dynamical age of 1.6$\times10^{6}$ yr for \HII region Sh2-104.

2. Five large clumps are found to be regularly distributed along the molecular ring. One of these clumps is associated with an $\uchii$ region and IRAS 20160+3636. The column density and mass of the molecular ring is 6.8$\times10^{21}$ cm$^{-2}$ and $\sim$2.2$\times10^{4}\rm M_{ \odot}$. For the CO molecular ring surrounded \HII region Sh2-104, we also did not detect the CO emission of higher than 3 signal inside the ring. The north-east portion of the ring is blueshifted relative to the LSR velocity of 1 $\kms$, while the south-west portion is redshifted. Hence, the ring is expanding.   We concluded that the collected gas emission around \HII region Sh2-104 may be a two-dimensional ring-like structure. 

3.  From the high-resolution (18$\farcs2$) column density map constructed by the Hi-GAL survey data, we extract 21 clumps. The mass-radius relationship for the clumps shows that about 90\% of all the identified clumps will form low-mass stars. Moreover, we evaluate the  stability of these clumps against gravitational collapse using the virial parameter, finding only three clumps collapse. A power-law fit to the identified clumps yields $M=281M_{\odot}(r/\rm pc)^{1.31\pm0.08}$. The slope is  similar to what was found in \citet{Kauffmann2010}.

4. The selected YSOs  are distributed along the ring around \HII region Sh2-104.  Comparing the dynamical age of \HII region Sh2-104 with the YSOs timescale and the fragmentation time of the molecular ring, we further conclude that these YSOs may be triggered by \HII region Sh2-104, and the CC process is at work in this region.

\acknowledgments We thank the referee for his/her report which helps to improve the quality of the paper.  We are also grateful to the staff at the Qinghai Station of PMO for their assistance during the observations.  This work was supported by the National Natural Science Foundation of China (Grant No. 11363004, 11403042, 11673066, and 11703040).

\end{document}